\documentclass[10pt]{article}
\usepackage[left=0.8in,top=0.8in,right=0.8in,bottom=0.8in]{geometry}

\usepackage{graphicx}
\usepackage{amsmath}
\usepackage{amssymb}
\usepackage{amsthm}
\usepackage{latexsym}
\usepackage{bm}
\usepackage{array,supertabular}
\usepackage{color}
\usepackage{framed}
\usepackage{setspace}
\usepackage{fancyhdr}
\usepackage{stmaryrd}
\usepackage{mathrsfs}
\usepackage{mdframed}
\usepackage{url}
\usepackage{multirow}
\usepackage{algorithm}
\usepackage{algpseudocode}
\usepackage{subcaption}
\usepackage{makecell}

\newcolumntype{P}[1]{>{\centering\arraybackslash}p{#1}}

\SetSymbolFont{stmry}{bold}{U}{stmry}{m}{n}

\numberwithin{equation}{section}

\begin{document}
\title{Large-eddy simulation of the FDA benchmark blood pump: validation against experiments and implications for turbulent flow mechanisms}
\author{Xuanming Huang$^{\textup{ a}}$, Chi Ding$^{\textup{ a}}$, Yujie Sun$^{\textup{ a}}$, Shidi Huang$^{\textup{ a}}$, Andrea Cioncolini$^{\textup{ b}}$, \\ Damiano Padovani$^{\textup{ b}}$, Ju Liu$^{\textup{ a}, *}$\\
$^a$ \textit{\small Department of Mechanics and Aerospace Engineering,}\\
\textit{\small Southern University of Science and Technology,}\\
\textit{\small 1088 Xueyuan Avenue, Shenzhen, Guangdong 518055, China}\\
$^b$ \textit{\small Department of Mechanical Engineering (Robotics),}\\
\textit{\small Guangdong Technion-Israel Institute of Technology,}\\
\textit{\small Shantou, Guangdong 515063, China}\\
$^{*}$ \small \textit{E-mail address:} liuj36@sustech.edu.cn}
\date{}
\maketitle

\section*{Abstract}
This study presents a systematic validation and comparative assessment of computational fluid dynamics (CFD) strategies for centrifugal blood pump simulations using the U.S. Food and Drug Administration benchmark model. A scale-resolving large eddy simulation (LES) with transient sliding-interface (SI) coupling is evaluated and compared against Reynolds-averaged Navier-Stokes (RANS) approaches employing both multiple reference frame and SI formulations. Numerical predictions are validated through direct comparison with particle image velocimetry measurements under two representative operating conditions. The results indicate that LES with transient rotor-stator coupling achieves consistently improved agreement with experimental velocity fields compared with RANS-based methods, particularly in the diffuser region where strong intermittency and wall-bounded turbulence are present. In contrast, RANS-based approaches exhibit noticeable discrepancies in these regions. A mesh sensitivity study and an assessment of temporal averaging effects are conducted for LES. The quality of the LES results is further quantified using three complementary metrics, demonstrating that a mesh resolution of approximately 80 million cells achieves a well-resolved LES regime. Building on the validated scale-resolving simulations, detailed analyses of vortical structures, turbulent kinetic energy distributions, and velocity energy spectra are performed to characterize the internal flow physics of the pump. The results reveal the central role of transient vortex dynamics in turbulence generation within centrifugal blood pumps. Overall, this study demonstrates that scale-resolving, transient simulation approaches are essential for accurately capturing the highly unsteady, turbulence-dominated flow features in ventricular assist devices and provides practical guidance for future high-fidelity hemodynamic and hemocompatibility studies.

\vspace{5mm}

\noindent \textbf{Keywords:} Ventricular assist device, Centrifugal pump, Turbulence, Computational fluid dynamics, Rotor-stator coupling, Validation

\section{Introduction}
Heart failure demands prompt and effective treatment to improve patient outcomes and quality of life. Ventricular assist devices (VADs) have emerged as a vital therapeutic option, providing mechanical circulatory support either as a bridge to heart transplantation or as destination therapy for patients ineligible for organ transplantation. Owing to the persistent shortage of donor organs, VADs have drawn increasing clinical and research interest. First-generation VADs use membrane- or diaphragm-based displacement pumping to produce pulsatile flow and continue to be widely used in pediatric patients. In contrast, rotary pumps deliver continuous blood flow and offer a more compact and biocompatible design, leading to improved patient survival rates and widespread clinical adoption \cite{Loor2012}. Classical continuous-flow VADs primarily employ axial-flow pump configurations \cite{Miyamoto2022}. More recently, centrifugal pumps with magnetically levitated rotors have emerged as the third-generation designs \cite{Hoshi2006}, exemplified by devices such as the HeartMate III and HeartWare HVAD. By minimizing mechanical contact, these devices substantially reduce device wear and shear-induced blood damage, as non-physiological shear stresses are believed to damage blood cells membrane and intracellular structures \cite{Chen2019}. In addition, modern extracorporeal membrane oxygenation systems typically employ centrifugal blood pumps based on similar operating principles, but with larger flow passage and lower head pressure compared with implantable VADs \cite{OBrien2017}.

Centrifugal VADs differ fundamentally from axial rotary VADs in both geometric scale and operating conditions. For the two aforementioned centrifugal VADs, their rotor diameter typically ranges between 40 and 60 mm, with flow rates of 3-7 L/min, and rotation speeds of 2000-7000 rpm under near-optimal operating conditions. In contrast, axial-flow VADs typically have a diameter below 20 mm, operate at higher speeds, ranging from 6000 to 45000 rpm, with flow rates of 1.5-6 L/min, within their clinically relevant operating range. Owing to their larger flow passages, centrifugal pumps can achieve the required hydraulic performance at substantially lower rotational speeds \cite{Fraser2011}, leading to flow conditions that differ markedly from those in axial-flow designs. These differences are reflected in the characteristic pump Reynolds-number regimes. Under typical operating conditions, the Reynolds number for axial VADs generally lies within the range of $10^4$-$10^5$, whereas that for centrifugal VAD spans $10^5$-$10^6$ \cite{Fraser2011}. The onset of turbulence in pumps is considered to occur when the pump Reynolds number exceeds $10^6$ \cite{WuLetter}. Consequently, centrifugal blood pumps often operate close to the nominal turbulence threshold, where laminar, transitional, and turbulent flow structures may simultaneously develop in different regions of the device, giving rise to complex internal flows. In axial-flow blood pumps, such complexity in flow patterns has been explicitly demonstrated. Torner et al. employed high-fidelity simulations to reveal pronounced turbulence production associated with secondary flows and vortical structures arising from interactions between the flow and devices \cite{Torner2018}. By contrast, the substantial differences in Reynolds-number range, geometric scale, and dominant flow mechanisms limit the direct applicability of these findings to centrifugal blood pumps, leaving their turbulence characteristics less well understood. A detailed characterization of the turbulent flow structures provides fundamental insights that guide the iterative design and optimization of blood pumps for clinical use.

To assess the accuracy of computational fluid dynamics (CFD) in flow prediction in blood-contacting medical devices, the U.S. Food and Drug Administration (FDA) established two benchmark problems: a nozzle model and a centrifugal pump model \cite{FDAbenchmark}. For each benchmark, experiments were conducted independently by three laboratories under controlled conditions, and corresponding numerical results were collected by the FDA for systematic comparison. The particle image velocimetry (PIV) data for the blood pump benchmark was first reported by Hariharan et al. \cite{Hariharan2018}, followed by a detailed comparison with computational modeling by Malinauskas et al. \cite{Malinauskas2017}. Despite extensive efforts, to our knowledge, no CFD study has yet been able to accurately reproduce the velocity consistently across operating conditions, especially in the diffuser region. As a result, reliable prediction of the velocity field therefore remains a significant challenge for computational investigations. This limitation further undermines the reliability of CFD-based blood damage estimations, which are known to be highly sensitive to the flow field predictions.

Many studies of the FDA benchmark blood pump have used the Reynolds-Averaged Navier-Stokes (RANS) equations as the modeling approach. While this approach captures the overall flow behavior at relatively low cost, it does not resolve instantaneous turbulent information. In contrast, large eddy simulation (LES) resolves the unsteady flow field by directly capturing the energy-containing turbulent structures and therefore provides a more detailed picture of the unsteadiness of the flow. However, owing to its higher computational cost, it has been applied in only a limited number of studies of the FDA benchmark pump. In addition to the choice of turbulence modeling strategy, modeling of blood pumps also differs in their treatment of rotor motion, which can be broadly classified as steady or transient. Steady approaches solve the flow in separate reference frames and efficiently capture mean rotational effects. Yet, such approaches neglect the unsteady interaction between the two subdomains. In contrast, transient methods explicitly account for the physical rotation of the rotor and enable time-resolved coupling between domains, thereby allowing more detailed resolution of unsteady flow features.

Ponnaluri et al. \cite{Ponnaluri2023} conducted a comprehensive assessment based on the FDA blood pump benchmark by compiling CFD results submitted by multiple participating groups using a range of numerical models. Although reasonable accuracy was achieved under individual operating conditions, no single approach consistently produced accurate predictions across all quantities and conditions. The variability observed among the submitted results highlights the strong sensitivity of CFD predictions to modeling choices and underscores the need for systematic validation to identify factors governing predictive relability. A number of independent studies have focused on assessing specific CFD methodologies and turbulence treatments. Good et al. \cite{Good2020} employed the $k$-$\omega$ shear-stress transport (SST) RANS model to simulate operating conditions 1, 4, and 5\footnote{The operating conditions follow the FDA benchmark specifications and are detailed in Table \ref{tab:operating conditions}.} with three different spatial resolutions. While their simulations capture the qualitative trends of velocity in the diffuser for conditions 4 and 5, the velocity magnitudes were consistently overestimated, and condition 1 remained difficult to predict accurately. Gross-Hardt et al. \cite{Gross2019} investigated the influence of turbulence model selection by comparing the $k$-$\omega$ SST model, the Stress-Blended Eddy Simulation (SBES) model, and the laminar model. Their findings indicated that the SBES model yielded the most accurate diffuser region velocity predictions, suggesting the importance of scale-resolving approaches. Semenzin et al. \cite{Semenzin2021} further evaluated several modeling strategies, including transient and steady methods combined with laminar, $k$-$\omega$ SST, and SBES models. They found that the steady approach for rotor-stator coupling produced unphysical velocity fields near the impeller tip, consistent with observations reported by Miccoli on shape optimization \cite{Miccoli2024}. In contrast, the transient approach combined with the SBES model showed improved agreement with PIV measurements. A limited number of investigations have explored LES for centrifugal blood pumps. Huo et al. \cite{Huo2021} performed LES of the FDA benchmark pump using temporal and spatial resolutions comparable to those of unsteady RANS. While the results outperformed RANS under identical resolutions, noticeable discrepancies remained in the diffuser region, highlighting the critical role of grid and temporal resolution in LES. Collectively, these studies indicate a growing trend toward the use of transient, scale-resolving approaches in the simulation of the FDA benchmark pump. Nevertheless, the majority of CFD studies on commercial blood pumps continue to rely on RANS-based approaches with relatively coarse meshes \cite{Gil2023, Han2022, Nissim2023}, consistent with the prevailing practices in FDA benchmark investigations. Despite the wide range of turbulence models and grid resolutions explored, no consensus has yet emerged regarding a reliable and optimal CFD strategy for blood pump simulations. This lack of agreement underscores the challenges inherent in modeling blood pumps and motivates further systematic validation studies. 

The present study aims to develop practical guidance for centrifugal blood pump simulation using a scale-resolving LES approach with transient rotor-stator coupling. Given the lack of consensus on optimal strategies for blood pump simulations, this work systematically evaluates the capability of LES to predict the flow field of the FDA benchmark pump through direct comparison with experimental measurements. The fidelity of LES predictions is assessed using three complementary quantitative quality metrics \cite{Celik2005, Gousseau2013, Klein2005}, which together provide robust support for the conclusions drawn regarding resolution requirements for reliable LES. Building on the validated scale-resolving results, the study further examines local flow features within the pump to elucidate the mechanisms governing the generation and evolution of turbulence in centrifugal blood pumps. These findings provide physically grounded insights into pump flow behavior and offer a useful reference for future high-fidelity numerical investigations.

\section{FDA centrifugal blood pump}
\begin{figure}[h!]
\centering
\begin{tabular}{cc}
\includegraphics[width=0.455\linewidth, trim=0 0 0 0, clip]{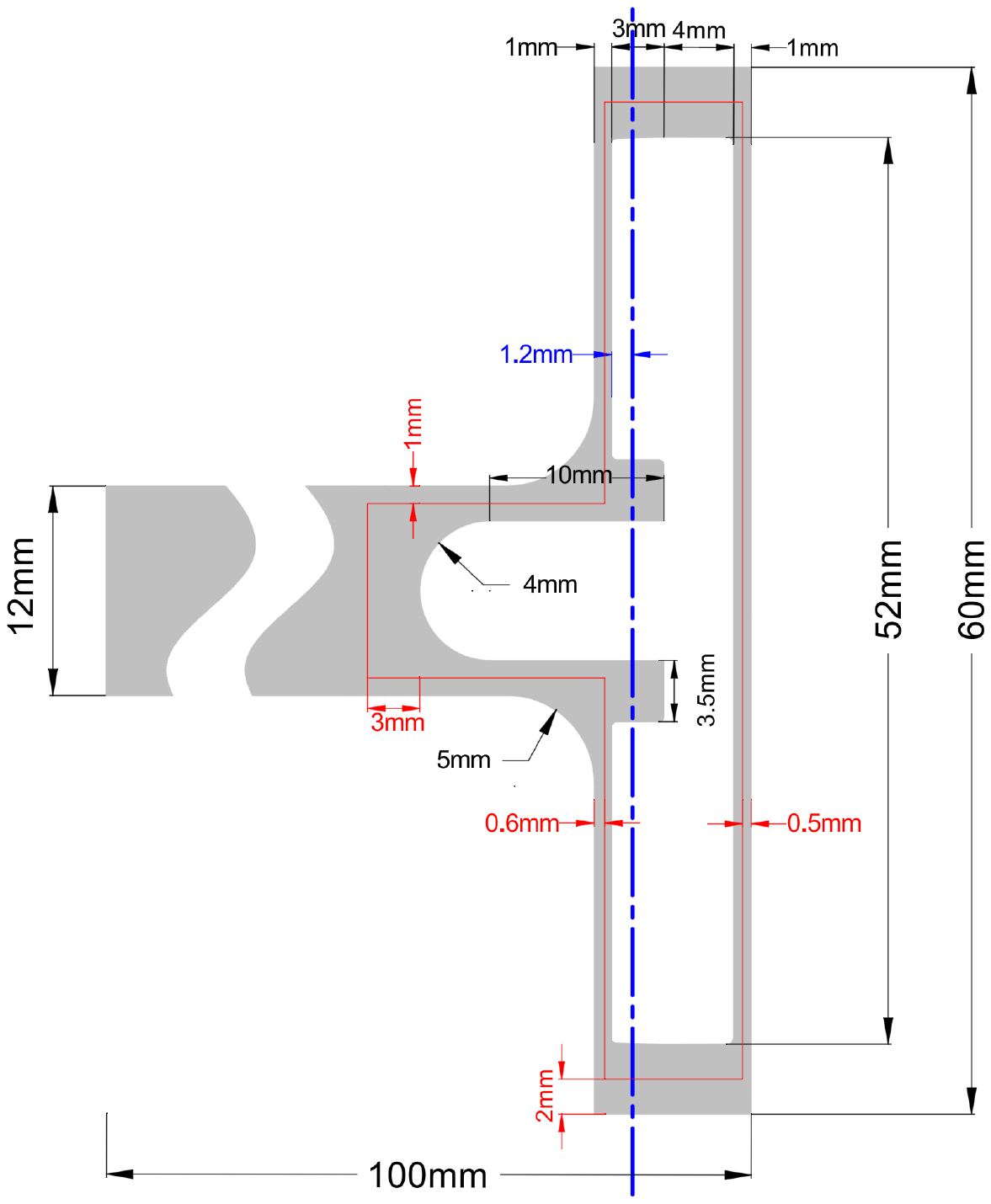} &
\includegraphics[width=0.455\linewidth, trim=0 0 0 0, clip]{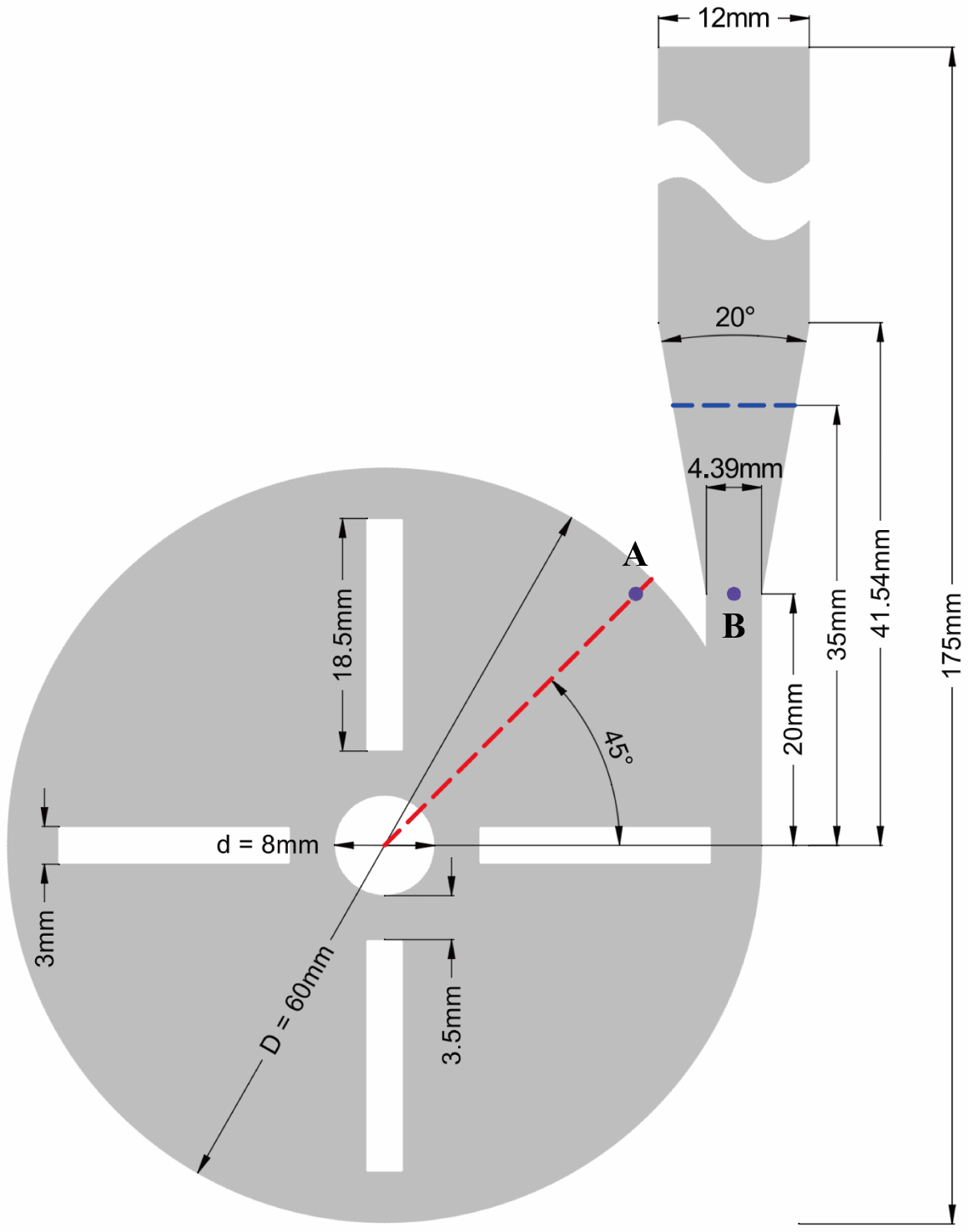} \\
\end{tabular}
\caption{The geometry of the FDA centrifugal blood pump. (a) The red line delineates the rotating subdomain in the computational model, and the blue line marks the sampling plane; (b) On this sampling plane, the red and blue dashed lines mark the locations for comparison with experimental data at the blade passage (the first quadrant) and the diffuser region, respectively. Points A and B denote the locations for velocity energy spectra analysis in Section 6.}
\label{fig:pump_geometry}
\end{figure}

In the FDA benchmark, the centrifugal pump was designed with a simple, well-defined geometry, consisting of a housing, rotor, inlet tube, and outlet nozzle. The rotor has four straight blades arranged at $90^{\circ}$ angles to one another, with all junctions between blade surfaces smoothly filleted. It is centrally positioned within the pump housing, with gap clearances of 4 mm radially. The inlet connects to a tube with a pressure transducer installed to record the pump inlet pressure. A blood analog fluid enters the pump through a central inlet, is accelerated by the rotor, and exits through a throat passage into an outlet diffuser. Detailed geometric specifications of the pump are illustrated in Figure \ref{fig:pump_geometry}, following the original geometric data reported in \cite{Malinauskas2017}.

The pump components were fabricated from optically clear acrylic to enable PIV measurements. Minor fabrication deviations were noted but remained within acceptable tolerances. The measurements were performed at multiple cross-sections, with a focus on flow features within the inflow region, blade passage, back-gap beneath the rotor, and the outlet diffuser region \cite{Hariharan2018}. In particular, detailed velocity fields were obtained on a plane located 1.2 mm below the top surface of the pump blade, focusing on flow features within the blade passage and the diffuser region. This plane, which is also used as the plane for data analysis in the present study, passes approximately through the geometric centerline of the throat and the outlet pipe, providing clear visualization of the key flow structures in these regions, as shown in Figure \ref{fig:pump_geometry}(b).

\begin{table}[H]
  \begin{center}
    \renewcommand{\arraystretch}{1.1}
    \begin{tabular}{ccccccc}
    \hline
    {Condition} & \makecell{Flow rate\\ $Q$({L/min})} & \makecell{Rotational speed\\ $n$({rpm})} & \makecell{$\mathrm{Re_{pump}}$} & \makecell{$\mathrm{Re_{pipe}}$} & \makecell{$\mathrm{Re_{throat}}$} & \makecell{Flow coefficient} \\
    \hline
    1  & $2.5$ & $2500$ & $209338$ & $1307$ & $3574$ & $0.00113$ \\
    2  & $2.5$ & $3500$ & $293073$ & $1307$ & $3574$ & $0.00081$ \\
    3  & $4.5$ & $3500$ & $293073$ & $2353$ & $6432$ & $0.00146$ \\
    4  & $6.0$ & $2500$ & $209338$ & $3138$ & $8577$ & $0.00272$ \\
    5  & $6.0$ & $3500$ & $293073$ & $3138$ & $8577$ & $0.00194$ \\
    6  & $7.0$ & $3500$ & $293073$ & $3661$ & $10006$ & $0.00226$ \\
    \hline
    \end{tabular}
  \end{center}
  \caption{Operating conditions of FDA benchmark centrifugal pump.
  The pump Reynolds number is defined as $\mathrm{Re_{pump}} = n D^2/\nu$, where $D$ is the impeller diameter, and $\nu$ is the kinematic viscosity. The Reynolds numbers for the inlet pipe and throat are defined as $\mathrm{Re} = 2 U R /\nu$ with $U = Q / \pi R^2$. Here, $U$ and $R$ stand for the local characteristic velocity and radius of the corresponding section. The dimensionless flow coefficient is defined as $Q/(n D^3)$.}
\label{tab:operating conditions}
\end{table}

The blood analog fluid used in the experiments is composed of sodium iodide, glycerin, and water. It behaves as a Newtonian fluid, and its refractive index matches that of the acrylic model, with only mild variations in viscosity and density. All experiments were conducted with the pump Reynolds number and flow coefficient maintained, and the results were rescaled using the dynamic viscosity $3.5\times 10^{-3}~\mathrm{Pa\cdot s}$ and density $1035~\mathrm{kg/m}^3$. Six operating conditions were tested in the interlaboratory experiments, as summarized in Table 1. Among these operating conditions, Condition 3 lacks measurements in the nozzle region, while Condition 4 exhibits the largest interlaboratory variability \cite{Hariharan2018}. Condition 5 is the most commonly used for validation \cite{Good2020, Huo2021, Ponnaluri2023, Semenzin2021}, as it corresponds to a high Reynolds number and moderate flow coefficient. Semenzin et al. demonstrated that variations in CFD settings led to noticeable differences in the predicted results under Condition 5 \cite{Semenzin2021}. Accordingly, it is adopted as the primary validation case. In addition, Condition 2 is also examined, which operates at the same high rotational speed as Condition 5 but corresponds to the lowest flow rate. 

\section{Methods}
\subsection{Models}
The flow in the blood pump is modeled by the filtered or averaged incompressible Navier-Stokes equations. Let $\bm u$ and $p$ denote the velocity and pressure, respectively. The governing equations can be written in the unified form
\begin{align*}
\frac{\partial \bar{\bm u}}{\partial t} + \nabla \cdot (\bar{\bm u} \otimes \bar{\bm u}) = - \frac{1}{\rho}\nabla {\bar{p}} + \nabla \cdot \bm \tau + \nabla \cdot \bm \tau_t + \bm f, \quad
\nabla \cdot \bar{\bm u} = 0,
\end{align*}
where the overbar denotes the filtered quantities in LES or ensemble-averaged quantities in RANS, $\rho$ is the fluid density, and $\bm f$ stands for the body force. The resolved viscous stress is defined as $\bm{\tau} = 2\nu \bar{\bm S}$, where $\nu$ is the kinematic viscosity and the components of the resolved strain-rate $\bar{\bm S}$ are given by $\bar{S}_{ij} :=(\partial\bar{u}_i/\partial x_j+\partial\bar{u}_j/\partial x_i)/2$. The additional stress $\bm \tau_t$ accounts for the effects of the turbulent motions and is provided by the turbulence closure model, such as the LES or RANS models described below.

\paragraph{LES closure}
In LES, the overbar denotes a spatial filtering operation that separates the resolved and subgrid scales. In this case, $\bm \tau_{t}$ is understood as the subgrid-scale stress and is modeled by $\bm{\tau}_t = 2 \nu_t \bar{\bm S}$, in which the eddy viscosity $\nu_t$ quantifies the effective diffusivity associated with the unresolved turbulent motions and accounts for the energy transfer between the resolved scales and the subgrid scales. In this study, $\nu_t$ is determined by the Wall-Adapting Local Eddy-viscosity (WALE) model \cite{Nicoud1999}, expressed as
\begin{align*}
\nu_t = l_s^2 \frac{(S^d_{ij}S^d_{ij})^{3/2}}{(\bar{S}_{ij}\bar{S}_{ij})^{5/2}+(S^d_{ij}S^d_{ij})^{5/4}}.
\end{align*}
In the above, $l_s = \min(\kappa d, C_w \Delta)$ is the filter length, $\kappa=0.41$ is the von K\'{a}rm\'{a}n constant, $d$ is the distance to the nearest wall, $C_w = 0.325$ is an empirical constant, and $\Delta$ is the local grid width, defined as the cubic root of the cell volume. The tensor $\boldsymbol{S}^d$ is the traceless symmetric part of the square of the filtered velocity gradient tensor, whose components are
\begin{align*}
S_{ij}^d = \frac{1}{2} (g_{ij}^2 + g_{ji}^2) - \frac{1}{3} \delta_{ij} g_{kk}^2, \quad  \quad g_{ij} = \frac{\partial \bar{u}_i}{\partial x_j},\quad \mbox{and} \quad g_{ij}^2 = g_{ik}g_{kj}.
\end{align*}
Although the use of LES for blood pump simulations is still relatively sparse, the WALE model is among the LES closures that have been previously employed for this benchmark configuration \cite{Crone2024,Wu2022}. It eliminates the need for near-wall damping functions and provides enhanced predictions of both wall-bounded and shear-dominated flows. Compared with the classical Smagorinsky model, it yields more accurate local dissipation near solid boundaries and correctly reproduces the near-wall scaling of the eddy viscosity. In the considered benchmark, the narrow clearance between the rotor and housing and the high rotation speed give rise to strongly shear-driven flow, rendering this model particularly well-suited. 

\paragraph{RANS closure}
In RANS, the overbar denotes an ensemble average, and $\bm \tau_t$ represents the Reynolds stress, accounting for the momentum transfer caused by the turbulent velocity fluctuations. Analogous to LES, the modeled stress is expressed as $\bm \tau_t = 2 \nu_t \bar{\bm S}$, where the turbulent viscosity $\nu_t$ represents the averaged turbulent fluctuations on the mean flow. In the classical RANS formulation, the governing equations are solved for a statistically steady mean flow, and all turbulent fluctuations are modeled. In contrast, unsteady RANS (URANS) retains the time dependence of the ensemble-averaged variables, allowing the resolution of large-scale, coherent unsteadiness in the mean flow, such as the blade-passing or geometry-induced periodic motions. In this study, the turbulent viscosity is obtained from the $k$-$\omega$ SST model, which solves two additional transport equations for the turbulent kinetic energy (TKE) $k$ and the specific dissipation rate $\omega$. This model combines the near-wall performance of the $k$-$\omega$ model with the free-stream robustness of the $k$-$\varepsilon$ model through a variable blending function. This hybrid approach has been widely validated for rotating and wall-bounded turbulent flows. Among the RANS closures assessed in CFD validation studies of the FDA blood pump benchmark, the $k$-$\omega$ SST model is considered a representative baseline option \cite{Good2020,Gross2019,Ponnaluri2023,Semenzin2021}.

\subsection{Treatment of stationary and rotating subdomains}
A major challenge in the numerical modeling of the blood pump lies in the treatment of the relative motion between the rotating impeller and the stationary housing. Two commonly used strategies for handling rotor-stator interaction are the multiple reference frame (MRF) method and sliding interface (SI) method. Both decompose the computational domain into a rotating subdomain associated with the impeller and a stationary subdomain associated with the housing. They differ fundamentally in how the rotation is modeled and how the two subdomains are coupled.

\paragraph{MRF method}
The MRF approach solves the governing equations in a reference frame that rotates with the impeller, whereas the flow away from the rotor region is solved in a stationary frame. Within the rotating region, additional inertial terms, most notably the Coriolis and centrifugal forces, arise from the transformation to a rotating frame. In the meantime, the computational mesh remains fixed throughout the entire domain. Owing to the absence of explicit time-dependent rotor-stator motion, the MRF method is most commonly employed with steady RANS models, providing an efficient steady-state approximation of rotating flows at a relatively low computational cost \cite{Malinauskas2017,Pauli2015,Ponnaluri2023}. A crucial aspect of the MRF approach is the placement of the interface between the rotating and stationary regions. To minimize the sensitivity of the solution to the interface position, it is generally recommended to place it within regions where the velocity gradients are relatively small \cite{Zadravec2007}. Yet, this is difficult to satisfy in blood pumps due to the extremely narrow tip-clearance space. As a result, a noticeable influence of the interface position on the predicted flow field has been observed in prior investigations \cite{Semenzin2021}.

\paragraph{SI method}
The SI method explicitly resolves the unsteady rotor-stator interaction by allowing relative motion between the rotating and stationary subdomains. The rotating region is handled within an arbitrary Lagrangian-Eulerian framework, in which the mesh undergoes a prescribed rigid-body rotation. In the meantime, the stationary subdomain is described in a conventional Eulerian framework. The two subdomains communicate through a non-conforming sliding interface, across which solution fields are exchanged at every time step to couple the rotor and stator. Flux transfer across the interface is typically constructed to preserve global conservation, often by evaluating the fluxes using the instantaneous geometric overlap between the moving and stationary control surfaces. Owing to its intrinsically time-dependent nature, the SI method is generally employed in conjunction with LES or URANS. Compared with the MRF approach, this method is better suited for capturing blade-passing effects and other strongly transient and periodic features. Although it is computationally more expensive due to mesh motion and temporal resolution, the SI method has been adopted in several high-fidelity studies of blood pumps \cite{Wu2022,Wu2024}.

\subsection{Mesh}
Three meshes, consisting of approximately 10 million, 51 million, and 80 million cells, respectively, were generated. Each mesh employs a hybrid discretization composed primarily of tetrahedral cells, supplemented by a 14-layer prismatic boundary-layer mesh adjacent to the walls, with a first-layer height of $6.5 \times 10^{-6}\ \rm{m}$ and a growth ratio of $1.1$. Mesh characteristics are summarized in Table \ref{tab:mesh}. Here, $y^+ := u_{\tau}y/\nu$, where $u_{\tau}$ is the friction velocity and $y$ is the distance from the first cell center to the wall. Maintaining $y^+ < 1$ ensures that the first off-wall grid point lies fully within the viscous sublayer, such that the near-wall velocity gradients are directly resolved. The maximum $y^+$ values for the three meshes remain close to $0.9$. As the mesh is refined, sharper near-wall gradients are better resolved, leading to a mild increase in the friction velocity and consequently a slightly higher maximum $y^+$ value on finer meshes. Overall, all three meshes achieve desirable sub-unit $y^+$ values, confirming adequate near-wall resolution for the present study.

\begin{table}[H]
  \centering
  \renewcommand{\arraystretch}{1.2}
  \setlength{\tabcolsep}{5pt}
  \begin{tabular}{c|cc|cc|c|c}
    \hline
    \multirow{2}{*}{Mesh} & 
    \multicolumn{2}{c|}{Cells in stationary domain} & 
    \multicolumn{2}{c|}{Cells in rotating domain} & 
    \multirow{2}{*}{Total} &
    \multirow{2}{*}{Maximum $y^+$} \\
    \cline{2-5}
     & Tets & Prisms & Tets & Prisms & \\
    \hline
    1 & 2,937,811 & 3,592,792 & 1,204,878 & 1,776,404 & 9,511,885 & 0.854 \\
    2 & 20,939,979 & 12,886,958 & 10,398,353 & 6,363,728 & 50,589,018 & 0.888 \\
    3 & 34,469,563 & 17,828,860 & 19,145,137 & 8,835,064 & 80,278,624 & 0.907 \\
    \hline
  \end{tabular}
  \caption{Summary of mesh resolution.}
  \label{tab:mesh}
\end{table}

\subsection{Computational setup}
The governing equations are solved using the finite volume method. The fluid was assumed to be Newtonian, with material properties matched to those used in the experiments (density of $1035~\mathrm{kg/m}^3$ and a dynamic viscosity of $3.5\times 10^{-3}~\mathrm{Pa\cdot s}$). The corresponding kinematic viscosity is $\nu = 3.382 \times 10^{-6}~\mathrm{m^2/s}$. Prior LES studies showed that the inlet length, inflow profile, and the inlet turbulent intensity have negligible effects on the pump performance prediction \cite{Wu2019,Xiang2023}, which is consistent with our own findings. We further verified that the outlet length can be shortened without affecting the computational results. Accordingly, to reduce computational cost, both the inlet and outlet tubes were truncated, yielding a final geometric model of 175 mm in length and 100 mm in height. No-slip boundary conditions were imposed on all housing and rotor surfaces. A zero-pressure condition was specified at the outlet, while a constant flow rate was prescribed at the inlet. The time step size corresponded to rotational increments of $0.6^\circ$, $0.3^\circ$, and $0.225^\circ$ for the 10M, 51M, and 80M meshes, respectively.

For the LES cases, the bounded second-order implicit scheme was employed for time marching, while the spatial discretization was carried out using the bounded central differencing scheme. These schemes provide near-uniform second-order accuracy over the majority of the computational domain. For the RANS simulations, the momentum equations were discretized using the second-order upwind scheme, and the URANS employed the same time stepping scheme as used in the LES. These numerical setups ensure a fair comparison between the different approaches. Each LES simulation was initialized from a URANS solution obtained after five rotor revolutions and was then continued for additional rotor revolutions to reach fully developed flow. Statistical averaging was performed at $90^\circ$ phase intervals for the SI approach, yielding four phase samples per rotor revolution. The phase-averaging procedure mirrors the experimental protocol, in which phase-averaged velocity fields were obtained from PIV measurements by averaging image pairs acquired with one blade positioned at $90^\circ$ relative to the pump outlet diffuser \cite{Hariharan2018,Ponnaluri2023}. All computations were executed on high-performance computing facilities. The most computationally demanding case, the LES using Mesh 3 for Condition 5, required approximately one day of wall-clock time per rotor revolution using 1080 Intel Xeon 8581 CPU cores.

\section{Validation}
In this section, we assess the predictive capability of LES and RANS approaches, together with different rotor-stator coupling strategies. All PIV data are reported with error bars representing the standard error, which quantify the uncertainty associated with independent measurements.

\begin{figure}[htbp]
  \centering
  \begin{tabular}{ccc}
    & \quad\quad {Blade passage}&
    \quad\quad {Diffuser} \\[0.5em]
    \raisebox{1.15\height}{\rotatebox{90}{Condition 2}} &
    \includegraphics[width=0.4\linewidth, trim=0 210 30 230, clip]{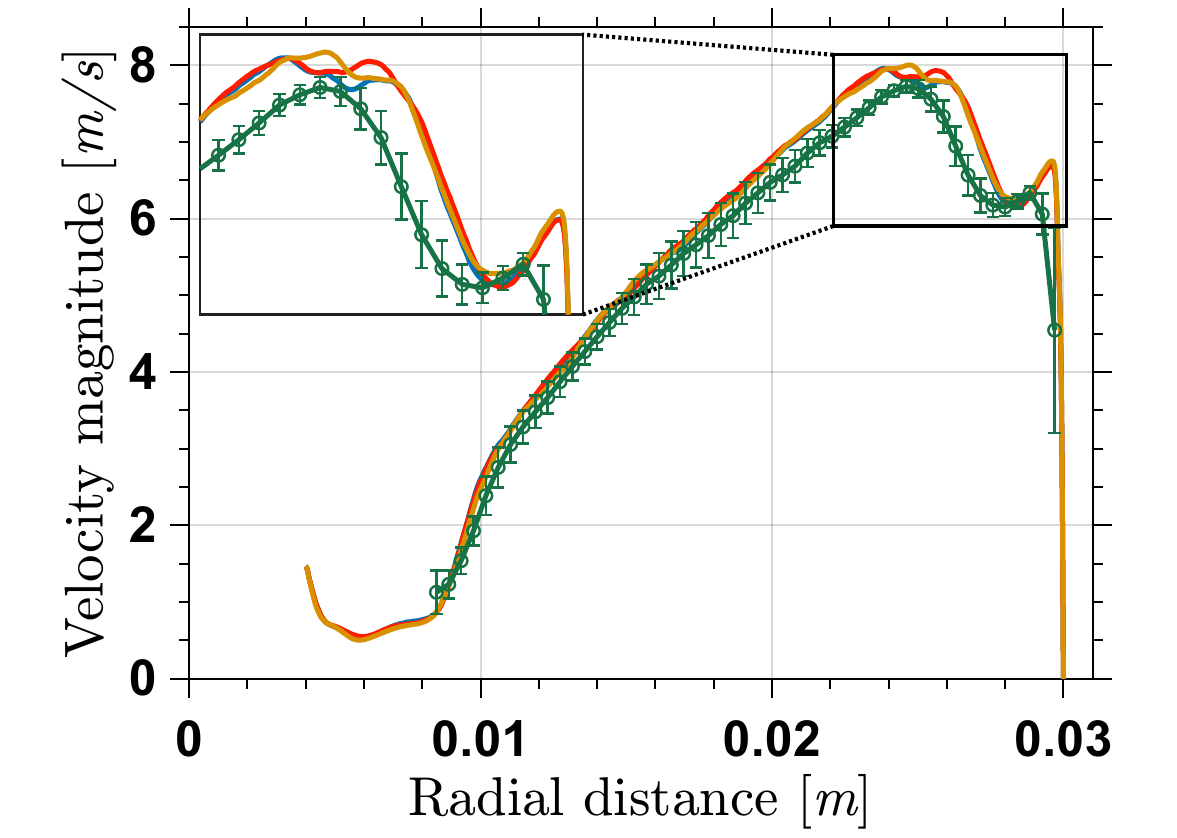} &
    \includegraphics[width=0.4\linewidth, trim=0 210 30 230, clip]{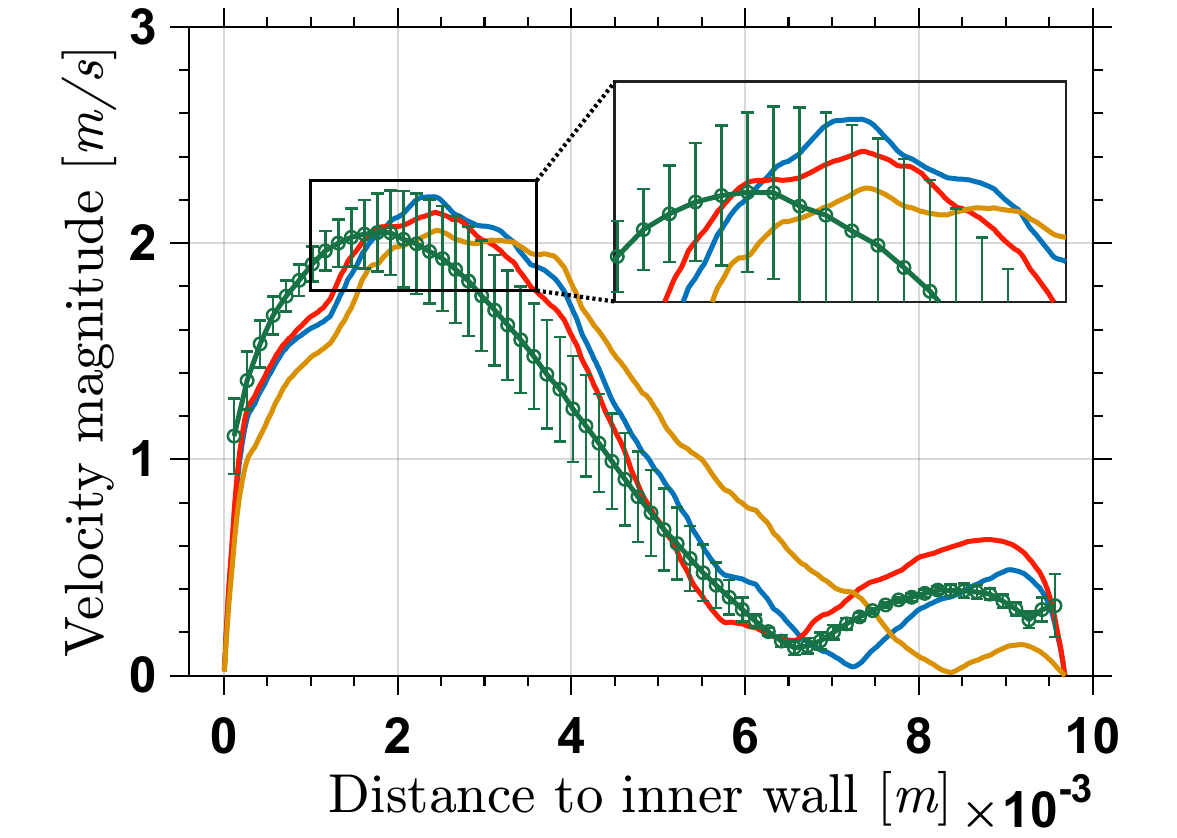} \\
    \raisebox{1.15\height}{\rotatebox{90}{Condition 5}} &
    \includegraphics[width=0.4\linewidth, trim=0 190 30 230, clip]{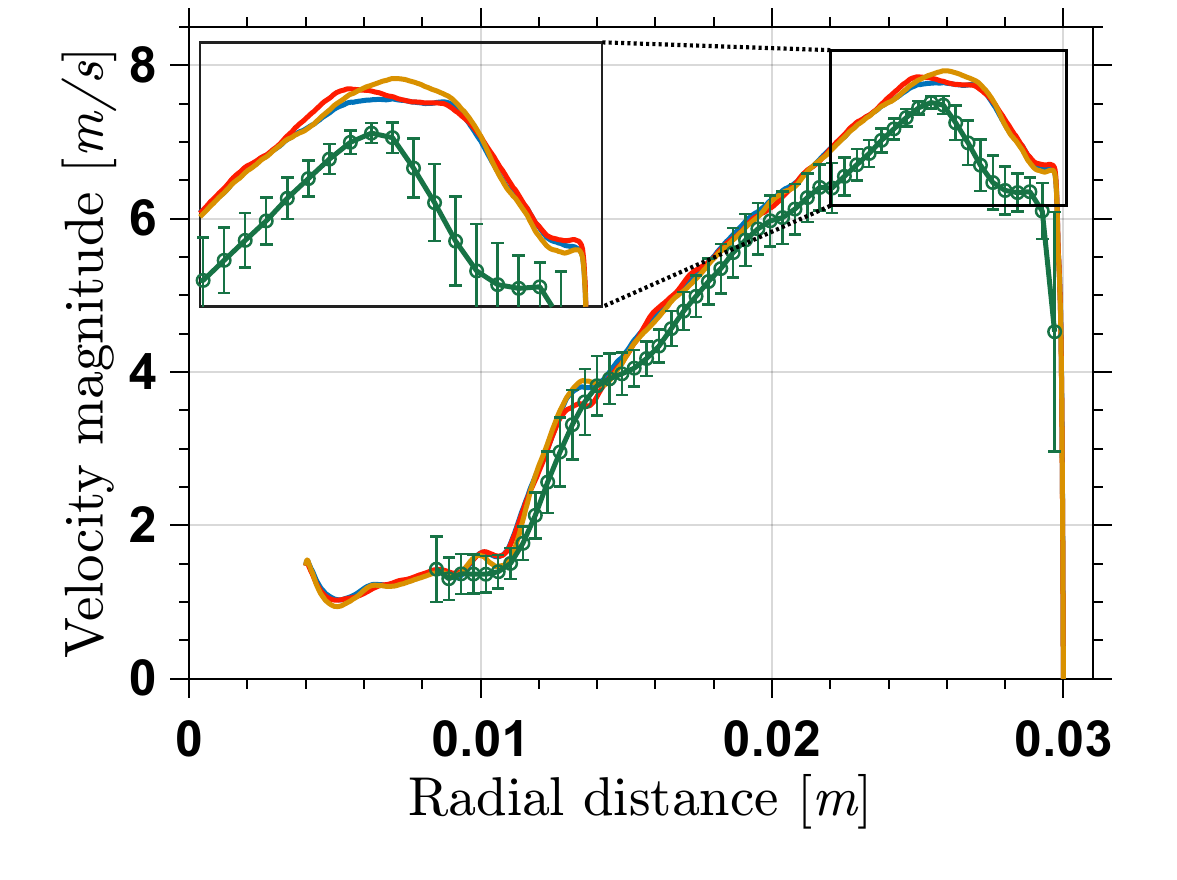} &
    \includegraphics[width=0.4\linewidth, trim=0 190 30 230, clip]{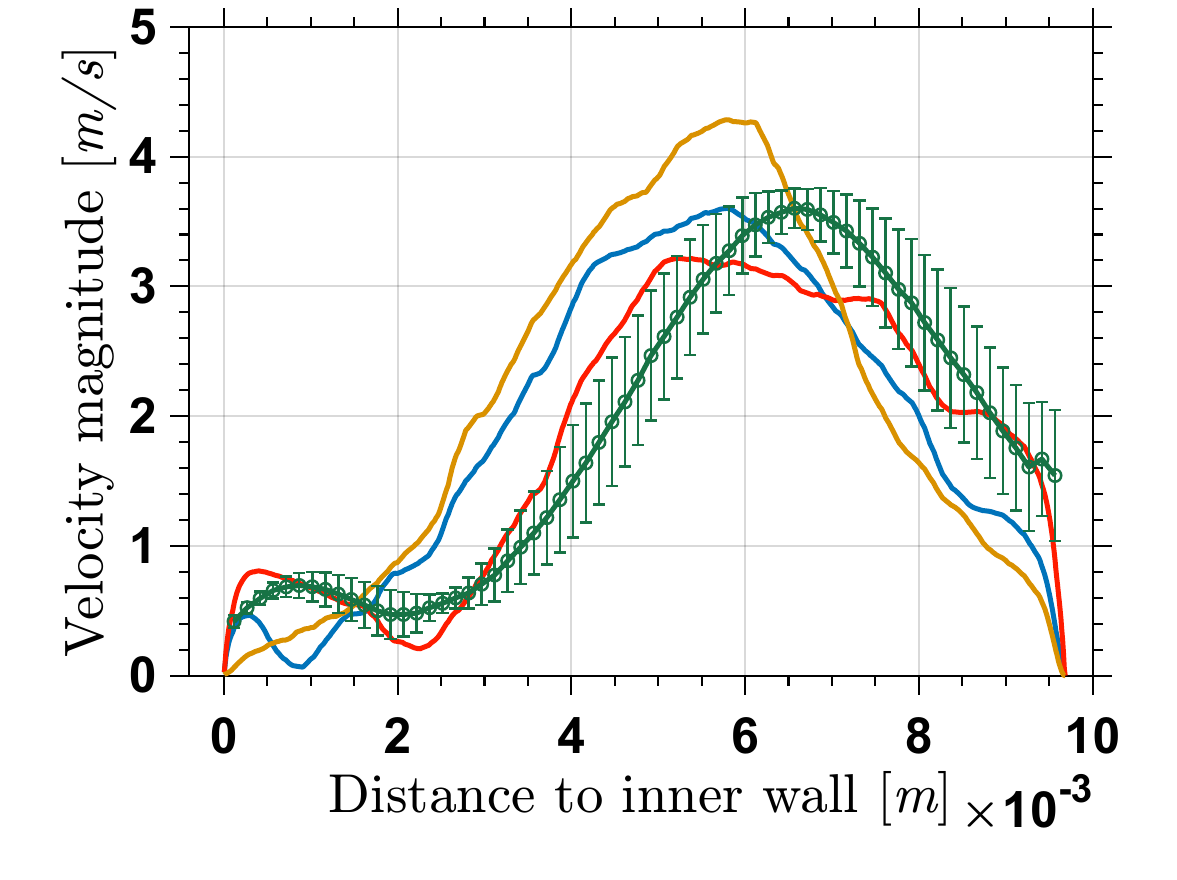} \\
  \end{tabular}
  \includegraphics[width=0.48\linewidth, trim=0 0 0 0, clip]{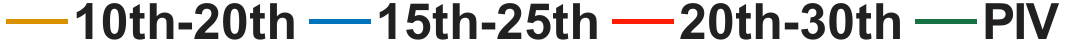}
  \caption{Effects of sampling interval on LES velocity profiles compared with PIV measurements in the blade passage and diffuser regions using Mesh 3. }
  \label{fig:80M_different_time}
\end{figure}

We first examine the time required for the flow to reach a statistically stationary state. Velocity fields in the blade passage and diffuser regions are collected from the 10th to the 30th rotor revolutions using Mesh 3. Figure \ref{fig:80M_different_time} compares the mean velocity profiles obtained from phase-averaged LES over successive 10-rotation sampling intervals with the corresponding PIV measurements. The results indicate that the velocity field in the blade passage converges rapidly for both conditions. In contrast, the flow in the diffuser requires a longer averaging period, particularly under Condition 5. For Condition 2, both the jet direction and the peak velocity magnitude attain statistical stationarity within the first sampling interval. On the opposite side of the primary peak, the velocity profile exhibits a decelerated region with a secondary local peak, whose convergence requires a longer averaging period. Based on these observations, all subsequent comparisons with other LES and URANS simulations are performed using velocity fields averaged over the 20th to 30th rotor revolutions.

\begin{figure}[htbp]./
  \centering
  \begin{tabular}{ccc}
    & \quad\quad {Blade passage}&
    \quad\quad {Diffuser} \\[0.5em]
    \raisebox{1.15\height}{\rotatebox{90}{Condition 2}} &
    \includegraphics[width=0.4\linewidth, trim=0 210 30 230, clip]{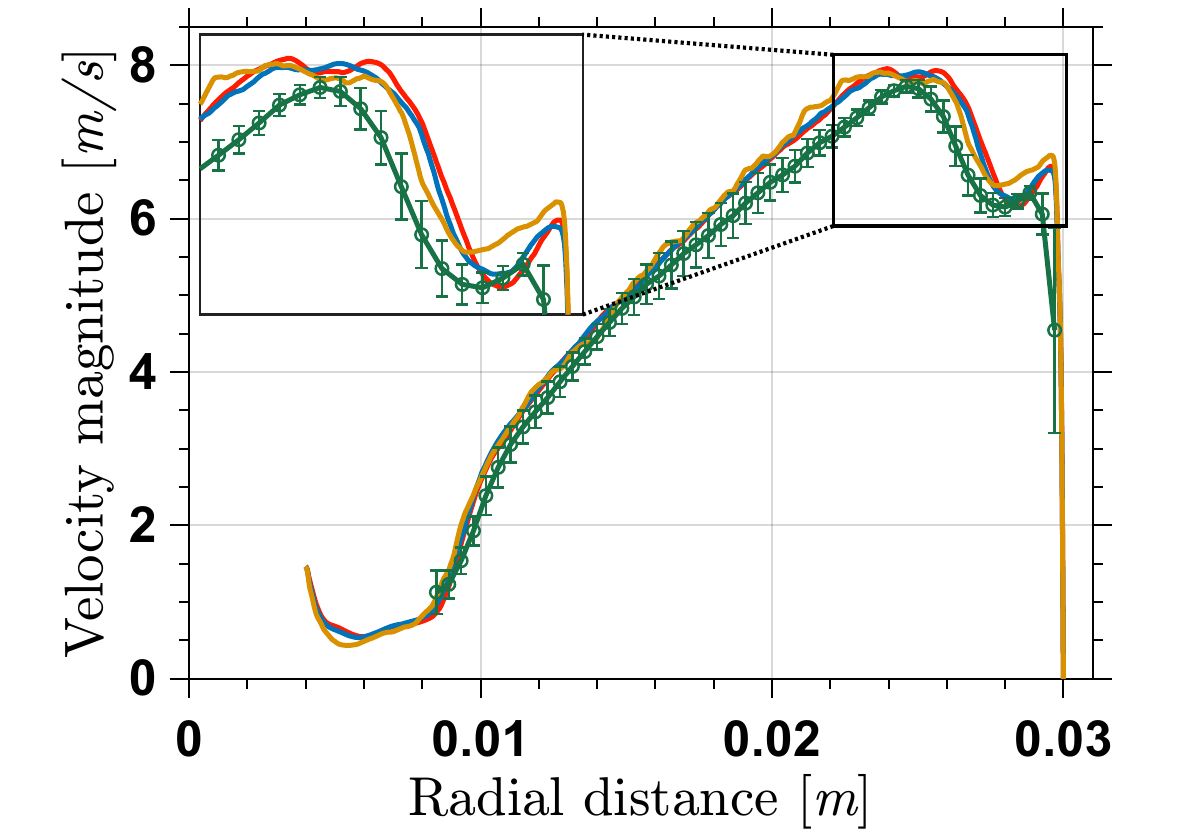} &
    \includegraphics[width=0.4\linewidth, trim=0 210 30 230, clip]{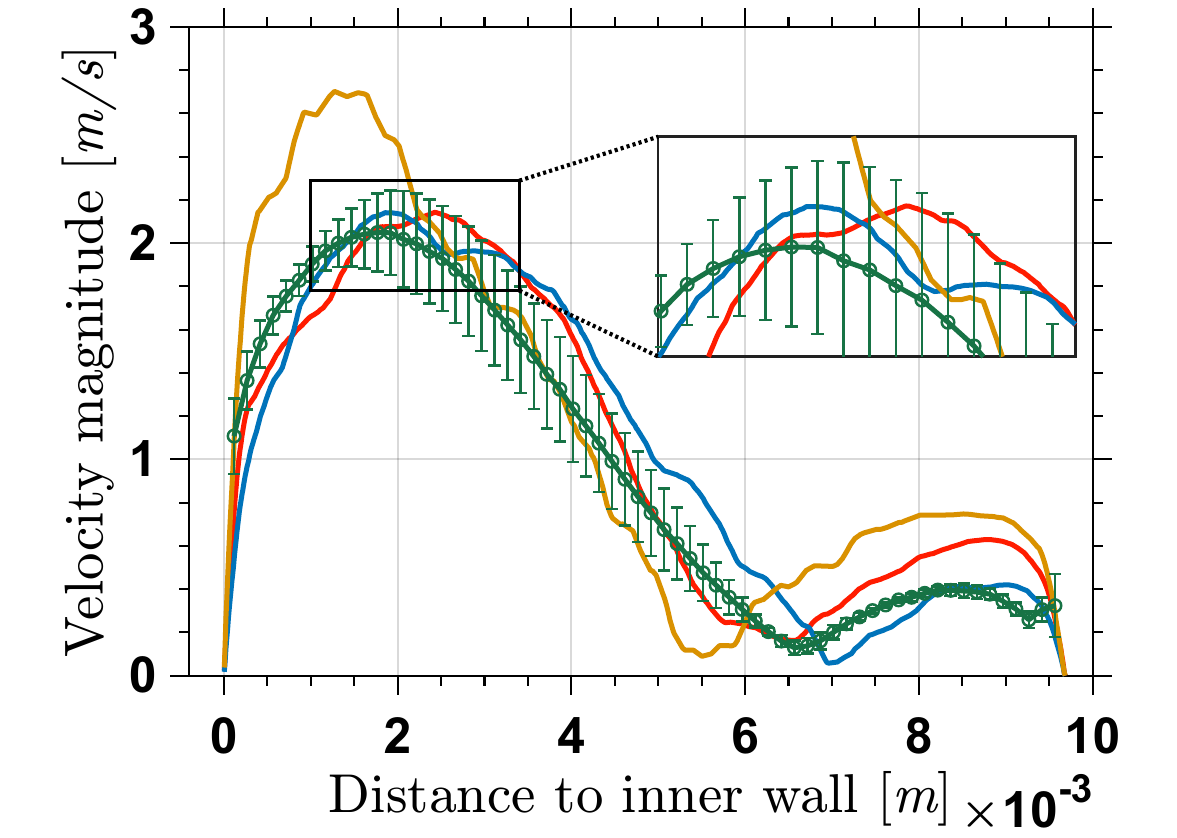} \\
    \raisebox{1.15\height}{\rotatebox{90}{Condition 5}} &
    \includegraphics[width=0.4\linewidth, trim=0 200 30 230, clip]{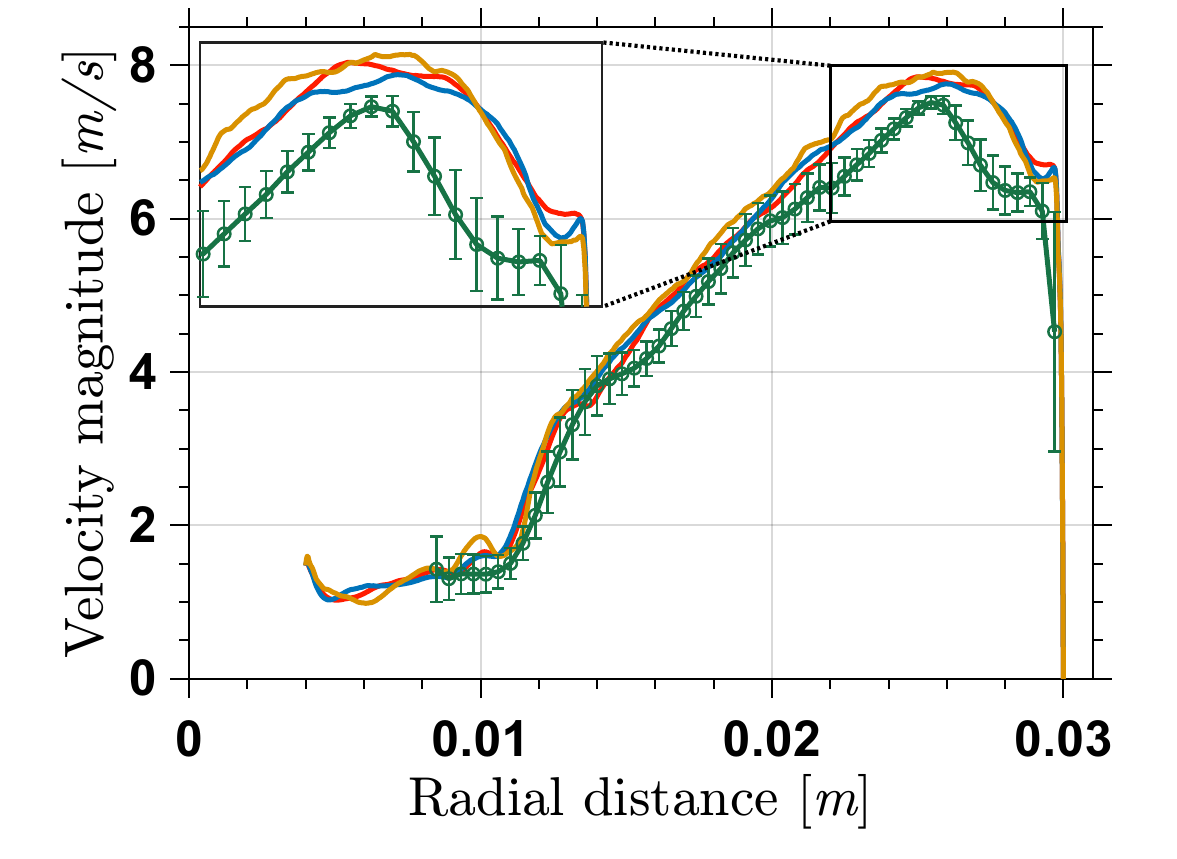} &
    \includegraphics[width=0.4\linewidth, trim=0 200 30 230, clip]{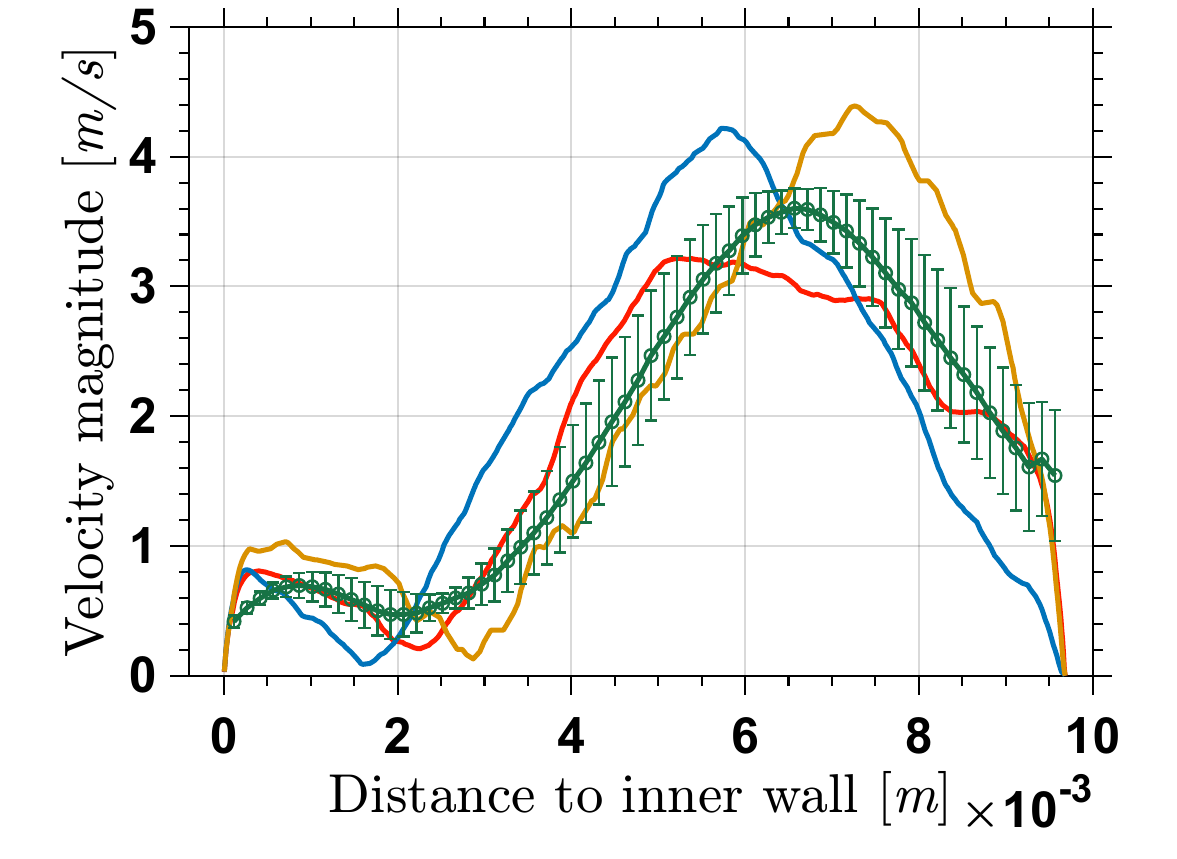} \\
  \end{tabular}
  \includegraphics[width=0.48\linewidth, trim=0 410 0 410, clip]{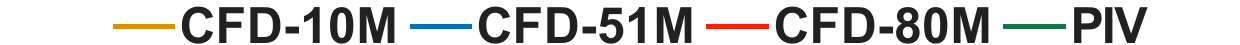}
  \caption{Velocity profile comparison between LES using three meshes and PIV experiments in the blade passage and diffuser regions.}
  \label{fig:mean_velo_20r_30r}
\end{figure}

Figure \ref{fig:mean_velo_20r_30r} compares the mean velocity profiles obtained from LES using different meshes and PIV measurements in the blade passage and diffuser regions. In the blade passage, the LES velocity profiles exhibit close agreement, indicating that the predicted mean velocity in this region is largely insensitive to spatial mesh resolution. Near the blade root, the numerical results closely match the experimental data. Toward the blade tip ($r \approx 26~\mathrm{mm}$, $r$ stands for the radial distance from the pump axis), the radial location of the peak velocity is consistently predicted, and all LES results exhibit a certain degree of overestimation compared with the measurements. In the clearance region ($26~\mathrm{mm} < r \leq 30~\mathrm{mm}$), all LES cases capture the pronounced velocity deceleration trend observed in the PIV measurements, and the profiles are consistently reproduced across all meshes. The strong consistency among the LES results suggests that the remaining discrepancies are unlikely to arise from numerical resolution or discretization effects and are more likely associated with experimental uncertainties and subtle differences between the simulations and experiments, including geometric variations and operating conditions.

In the diffuser region, the LES captures both the overall velocity distribution and the jet direction under both operating conditions. For Condition 2, the results obtained with Meshes 2 and 3 show close agreement with the experimentally measured velocity. In contrast, Mesh 1 significantly overestimates the peak velocity magnitude and predicts a jet peak location shifted toward the diffuser wall. In the central diffuser region, Mesh 3 provides the best overall agreement with the experimental data. However, in the deceleration zone near the opposite wall, Mesh 3 tends to overestimate the velocity, whereas Mesh 2 yields more accurate predictions. For Condition 5, both Mesh 1 and Mesh 2 overestimate the peak velocity magnitude and predict a jet peak location that deviates from the experimental measurements. Mesh 3 yields a closer prediction of the jet peak location, although it underestimates the peak velocity magnitude. Aside from the vicinity of the velocity peak, the predictions obtained with Mesh 3 closely agree with the experimental measurements over most of the diffuser region and largely fall within the PIV error bars. This represents a substantial improvement over previously reported results \cite{Semenzin2021}.

\begin{figure}[htbp]
	\centering
  \begin{tabular}{ccc}
	& \quad\quad {Blade passage}&
	\quad\quad {Diffuser} \\[0.5em]
	\raisebox{1.15\height}{\rotatebox{90}{Condition 2}} &
	\includegraphics[width=0.4\linewidth, trim=0 210 30 230, clip]{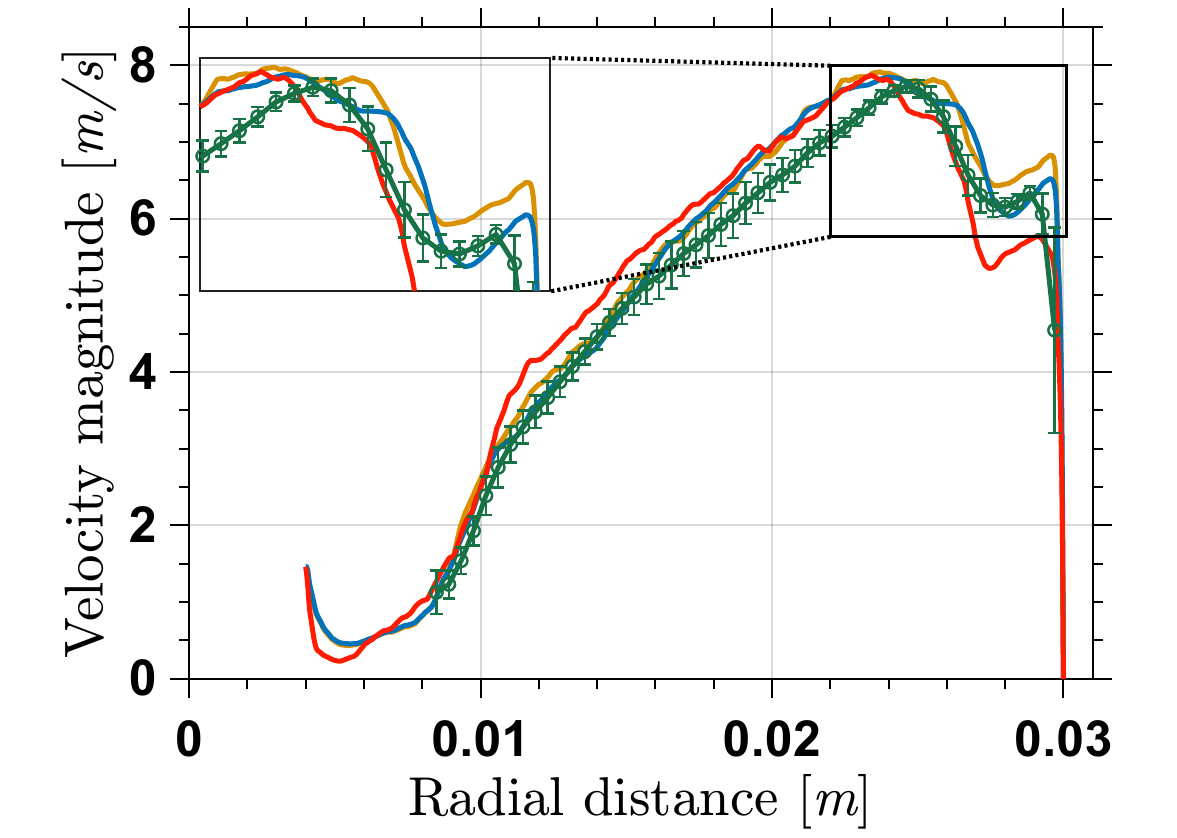} &
	\includegraphics[width=0.4\linewidth, trim=0 210 30 230, clip]{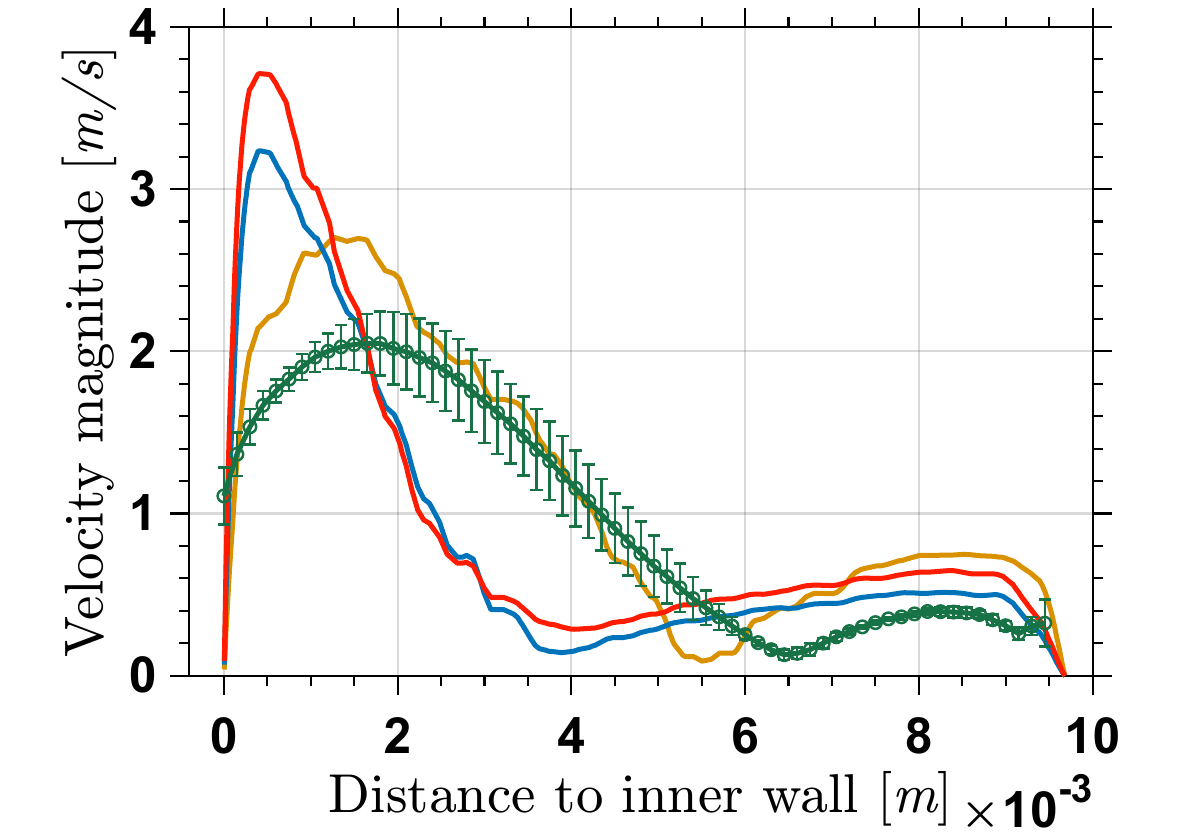}  \\
	\raisebox{1.15\height}{\rotatebox{90}{Condition 5}} &
	\includegraphics[width=0.4\linewidth, trim=0 200 30 230, clip]{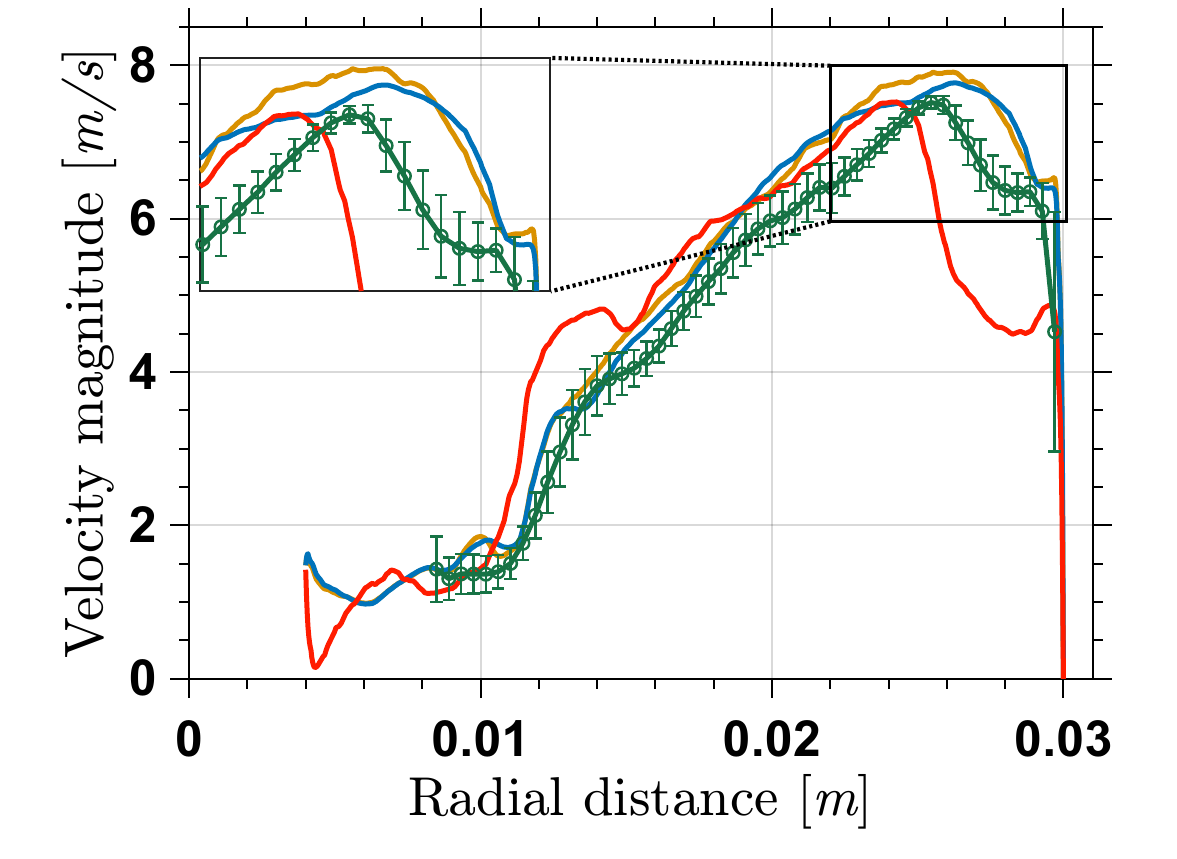} &
	\includegraphics[width=0.4\linewidth, trim=0 200 30 230, clip]{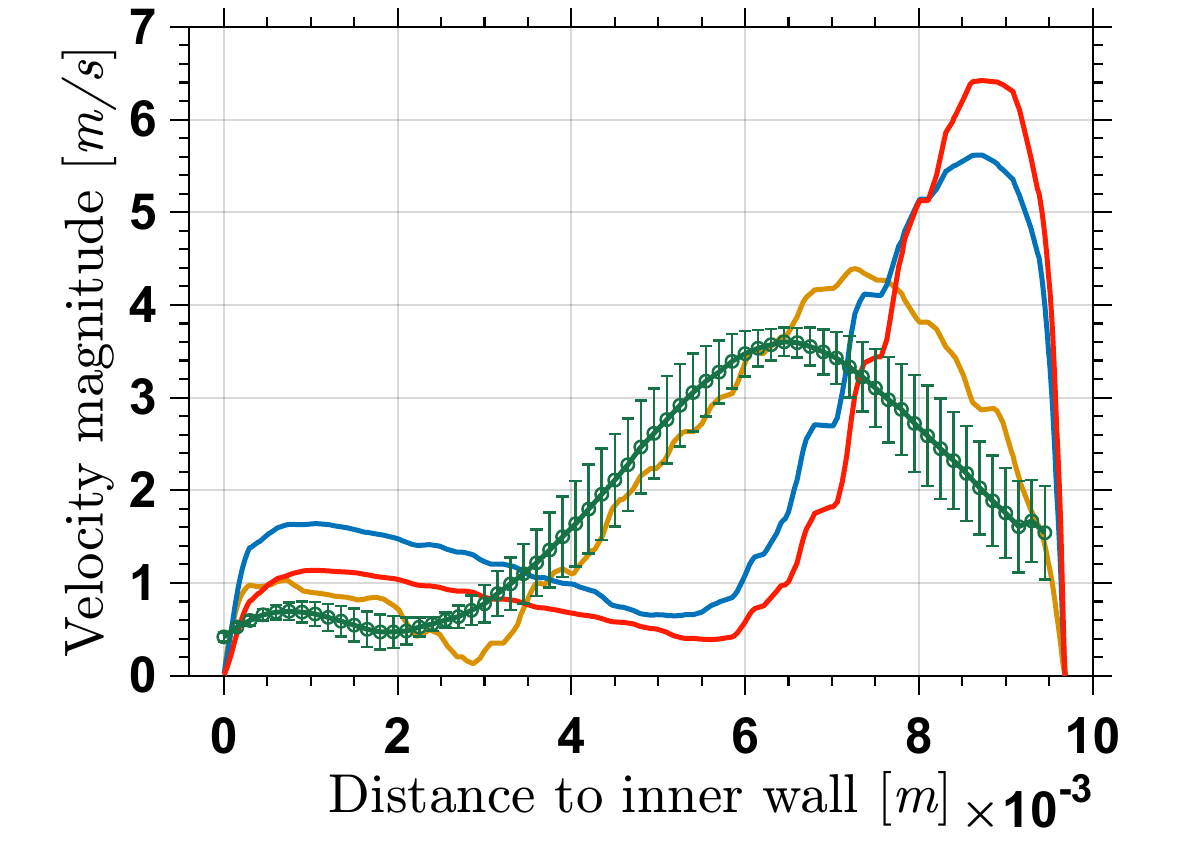} \\
\end{tabular}
\includegraphics[width=0.48\linewidth, trim=0 400 0 410, clip]{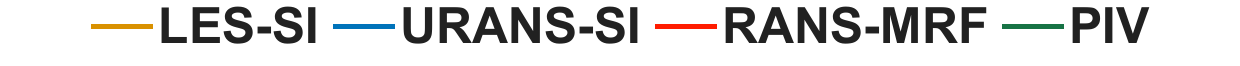}
\caption{Velocity profile comparison in the blade passage and diffuser regions between three modeling approaches and PIV experiments.}
\label{fig:mean_velo_20r_30r_methods}
\end{figure}

With the objective of providing guidance for the CFD analysis of blood pumps, the LES-SI approach is further compared with two widely used approaches: URANS with sliding interface (URANS-SI) and steady RANS with MRF (RANS-MRF). All three approaches are evaluated using Mesh 1, and both LES-SI and URANS-SI employ the same time-step size corresponding to a rotational increment of $0.6^{\circ}$. Figure \ref{fig:mean_velo_20r_30r_methods} presents the mean velocity profiles obtained using the three approaches under both operating conditions. In the blade passage region under Condition 2, LES-SI and URANS-SI produce similar velocity profiles, both exhibiting closer agreement with the experimental data than RANS-MRF. Under Condition 5, these discrepancies become more pronounced: RANS-MRF substantially underestimates the velocity in the tip clearance region ($26~\mathrm{mm} < r < 30~\mathrm{mm}$), and its predicted velocity profile exhibits noticeable spatial oscillations near the blade root ($4~\mathrm{mm} < r < 14~\mathrm{mm}$). In the diffuser region, although the LES-SI approach using Mesh 1 is less accurate than simulations employing finer mesh resolutions, it nonetheless represents a substantial improvement over the two RANS-based models under both operating conditions for jet prediction. Both URANS-SI and RANS-MRF significantly overestimate the peak velocity, and the jet peak locations are shifted toward the diffuser wall when compared to the experimental measurements. Notably, even the URANS formulation fails to adequately capture the flow profile in this region. These results demonstrate the improved predictive capability of LES-SI relative to RANS-based approaches for blood pump investigations.

\section{Assessment of LES quality}
In RANS, the solution typically converges to the statistically averaged flow field under grid refinement. In contrast, LES resolves only the large-scale turbulent motions and models the subgrid-scale ones using a filter whose characteristic width is related to the mesh size. As a result, the LES solution is inherently dependent on the grid resolution. The built-in uncertainty introduced by the filter length cannot be quantified through mesh refinement alone. This motivates the use of a posteriori metrics to assess the extent to which the turbulence is resolved and the simulation remains within the valid regime of LES modeling assumptions. In the following, three complementary metrics are applied to assess the quality of LES results for the considered benchmark pump. Throughout the analysis, $\langle \cdot \rangle$ stands for an appropriate statistical averaging operator used in the a posteriori analysis, which removes instantaneous fluctuations from the LES solution. We note that this averaging is distinct from the filtering operator (i.e. the overbar) employed in the LES modeling.

As the primary objective of LES is to directly resolve the energy-containing turbulent motions, Pope argued that a defining trait of an accurate LES is that the majority of the TKE is resolved by the simulation, and he suggested that at least $80\%$ of the total TKE should be resolved \cite{Pope2004}. This guideline is based on homogeneous isotropic turbulence and has since been widely invoked as a practical criterion. Celik et al. formalized Pope's concept into an LES quality index \cite{Celik2005} based on the assumption that the total TKE, denoted as $k^{\mathrm{tot}}$, can be decomposed into a resolved component $k^{\mathrm{res}}$ and an unresolved subgrid-scale component $k^{\mathrm{sgs}}$. The quality index, denoted by $\mathrm{IQ}_k$, is given as
\begin{align*}
\mathrm{IQ}_k := \frac{k^{\mathrm{res}}}{k^{\mathrm{tot}}} = \frac{k^{\mathrm{res}}}{k^{\mathrm{res}} + k^{\mathrm{sgs}}}.
\end{align*}
The resolved TKE is computed as 
\begin{align*}
k^{\mathrm{res}} = \frac12 \int_{\Omega}\langle {\mathbf{u}^{\prime}\cdot\mathbf{u}^{\prime}} \rangle dV,
\end{align*} 
where $\Omega$ denotes the entire pump domain, and $\mathbf{u}^{\prime} = \bar{\bm u} - \langle \bar{\bm u} \rangle$ is the velocity fluctuation. Values of $\mathrm{IQ}_k$ in the range of approximately $75\%$-$85\%$ are generally regarded as adequate for large-Reynolds-number flows in engineering problems. Sometimes, it is argued that resolving at least $90\%$ of the total TKE is necessary for reliable prediction of the mean flow fields \cite{Matheou2014}, and $\mathrm{IQ}_k$ values approaching $100\%$ correspond to DNS-like resolution quality \cite{Catellani2016}.

\begin{table}[H]
\centering
\renewcommand{\arraystretch}{1.2}
\setlength{\tabcolsep}{5pt}
\centering
\renewcommand{\arraystretch}{1.1}
\begin{tabular}{c|c|c|c|c|c}
\hline
& \multicolumn{1}{c|}{\multirow{2}{*}{\makecell[l]{Characteristic\\ mesh size $h$ (m)}}} & \multicolumn{2}{c}{Condition 2} & \multicolumn{2}{c}{Condition 5} \\
\cline{3-6}
\multicolumn{1}{c|}{} & & $k^{\mathrm{res}}(\mathrm{m^2/s^2})$ & $\mathrm{IQ}_k$ & $k^{\mathrm{res}}(\mathrm{m^2/s^2})$ & $\mathrm{IQ}_k$ \\
\hline
Mesh 1 & $1.572\times 10^{-4}$ & $4.434\times 10^{-6}$ & $80.48\%$ & $9.650\times 10^{-6}$ & $80.09\%$ \\
Mesh 2 & $9.004\times 10^{-5}$ & $5.018\times 10^{-6}$ & $91.09\%$ & $1.109\times 10^{-5}$ & $92.04\%$ \\
Mesh 3 & $7.719\times 10^{-5}$ & $5.250\times 10^{-6}$ & $95.29\%$ & $1.147\times 10^{-5}$ & $95.20\%$ \\
\hline
\end{tabular}
\caption{Resolved TKE $k^{\mathrm{res}}$ and corresponding $\mathrm{IQ}_k$ with $p = 2$.}
\label{tab:LES_IQ_k}
\end{table}

Since the subgrid-scale contribution $k^{\mathrm{sgs}}$ is not directly available, Celik et al. assumed that the unresolved TKE depends on the mesh resolution and follows a power-law of the form $k^{\mathrm{sgs}} = a_k h^p$, where $a_k$ is a coefficient, $h$ denotes a characteristic mesh size, and $p$ is the formal order of accuracy of the scheme. By performing two simulations of the same configuration using different meshes, a system of two equations can be constructed, from which $k^{\mathrm{tot}}$ can be extrapolated. For unstructured meshes, the mesh size $h$ can be defined in an averaged sense as $h = (\sum V_{\mathrm{cell}}/N_{\mathrm{cell}})^{1/3}$, where $V_{\mathrm{cell}}$ and $N_{\mathrm{cell}}$ denote the cell volume and the number of cells. In the present study, Meshes 1 and 3 are invoked to extrapolate $k^\mathrm{tot}$. The resolved TKE and the corresponding $\mathrm{IQ}_k$ values are summarized in Table \ref{tab:LES_IQ_k}. As expected, the LES quality improves with progressive mesh refinement. For Mesh 3, the values of $\mathrm{IQ}_k$ exceed $95\%$ for both conditions, suggesting that this resolution is sufficiently fine. For Mesh 2, the $\mathrm{IQ}_k$ values exceed $91\%$, which falls within the range generally regarded as adequate for engineering LES. For Mesh 1, the index values are around $80\%$, corresponding to the lower bound of acceptable LES resolution \cite{Celik2005}. This suggests the flow is likely only marginally resolved on Mesh 1.

\begin{figure}[htbp]
	\centering
	\begin{tabular}{cccc}
		\includegraphics[width=0.25\linewidth, trim=20 20 20 20, clip]{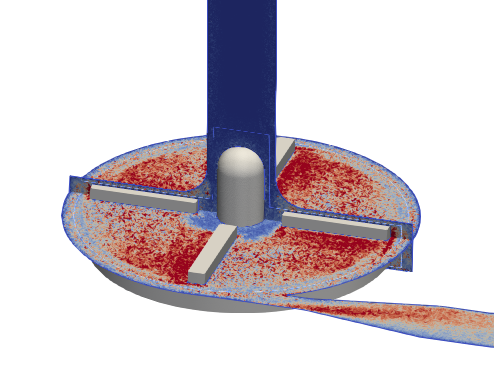} &
		\includegraphics[width=0.25\linewidth, trim=20 20 20 20, clip]{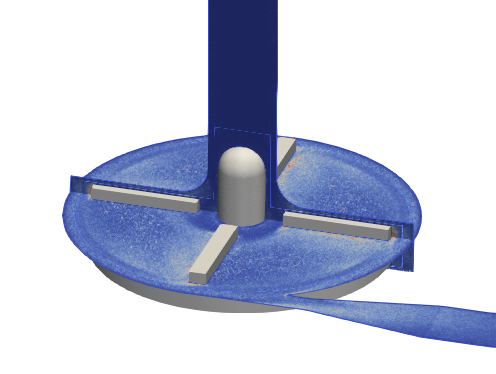}  &
		\includegraphics[width=0.25\linewidth, trim=20 20 20 20, clip]{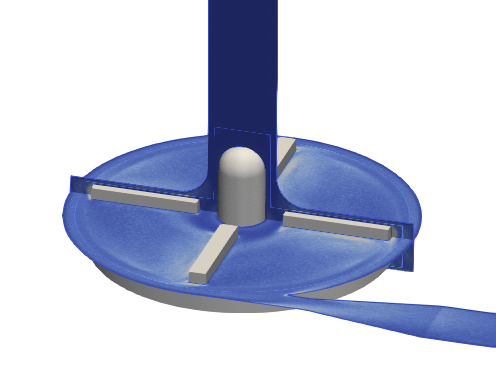} &
		\includegraphics[width=0.07\linewidth, trim=0 0 0 0, clip]{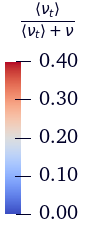} \\
		\includegraphics[width=0.25\linewidth, trim=20 20 20 20, clip]{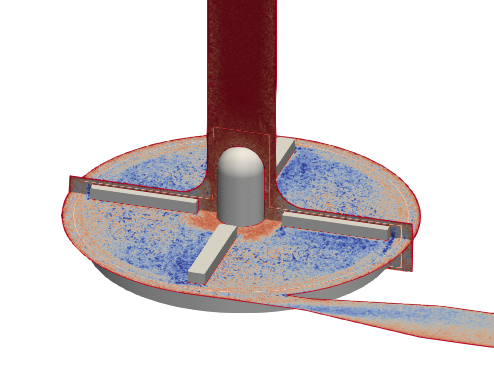} &
		\includegraphics[width=0.25\linewidth, trim=20 20 20 20, clip]{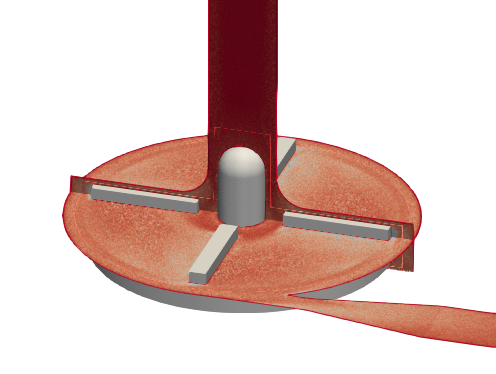}  &
		\includegraphics[width=0.25\linewidth, trim=20 20 20 20, clip]{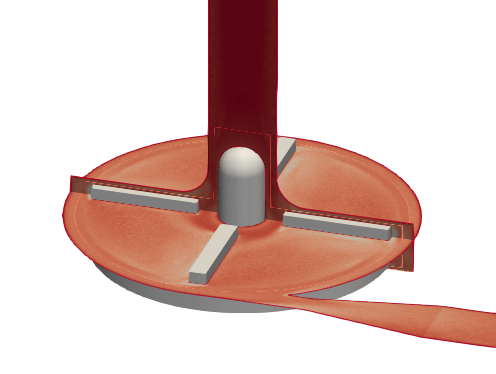} &
		\includegraphics[width=0.07\linewidth, trim=0 0 0 0, clip]{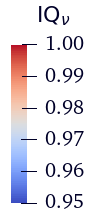} \\
		{Mesh 1} & {Mesh 2} & {Mesh 3} &
	\end{tabular}
	\caption{ Contours of subgrid activity parameter and $\mathrm{IQ}_{\nu}$ for three meshes under Condition 2.}
	\label{fig:contours_LES_quality_C2}
\end{figure}

\begin{figure}[htbp]
	\centering
	\begin{tabular}{cccc}
		\includegraphics[width=0.25\linewidth, trim=20 20 20 20, clip]{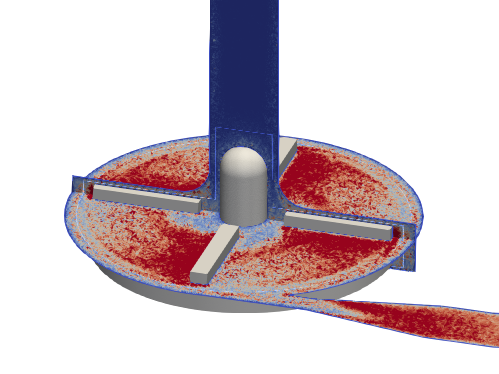} &
		\includegraphics[width=0.25\linewidth, trim=20 20 20 20, clip]{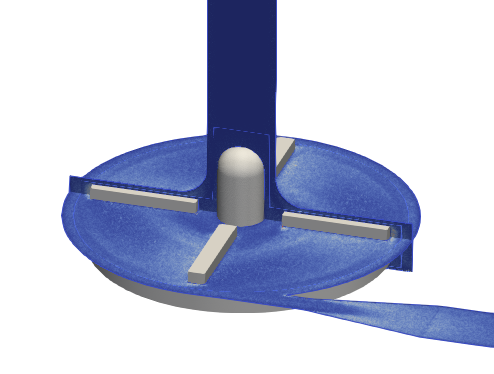}  &
		\includegraphics[width=0.25\linewidth, trim=20 20 20 20, clip]{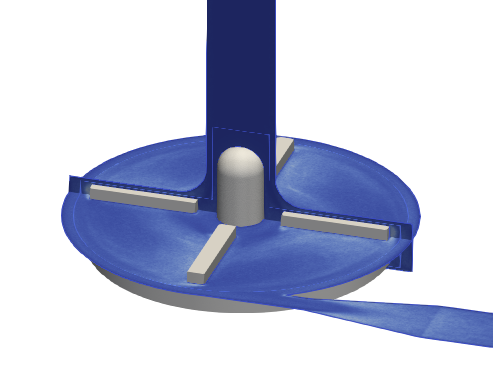} &
		\includegraphics[width=0.07\linewidth, trim=0 0 0 0, clip]{./legend_sub_act_para.png} \\
		\includegraphics[width=0.25\linewidth, trim=20 20 20 20, clip]{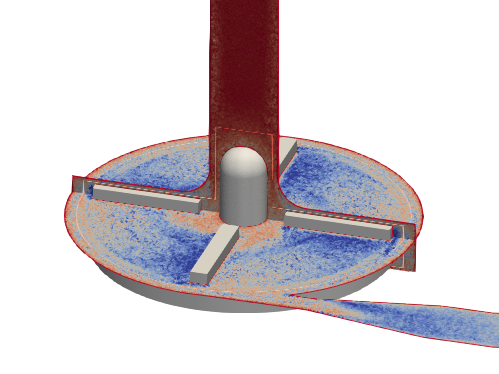} &
		\includegraphics[width=0.25\linewidth, trim=20 20 20 20, clip]{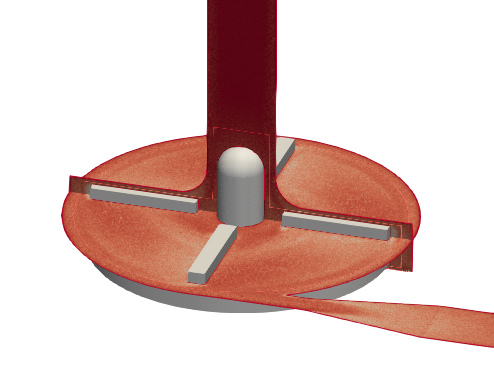}  &
		\includegraphics[width=0.25\linewidth, trim=20 20 20 20, clip]{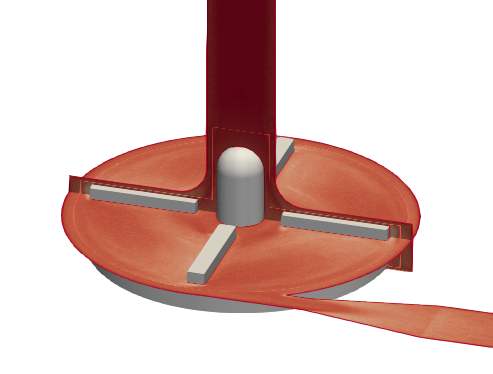} &
		\includegraphics[width=0.07\linewidth, trim=0 0 0 0, clip]{./legend_IQ_v.png} \\
		{Mesh 1} & {Mesh 2} & {Mesh 3} &
	\end{tabular}
	\caption{ Contours of subgrid activity parameter and $\mathrm{IQ}_{\nu}$ for three meshes under Condition 5.}
	\label{fig:contours_LES_quality_C5}
\end{figure}

While the resolved fraction of TKE provides a global measure of LES resolution, it does not explicitly account for how energy is transferred and dissipated across scales from a local perspective. Geurts and Fr{\"o}hlich proposed a metric based on the dissipation rates in LES \cite{Geurts2002}. They introduced the so-called subgrid activity parameter defined as $s := \langle\varepsilon_t\rangle/(\langle\varepsilon_t\rangle + \langle\varepsilon_{\mu}\rangle)$. The turbulent dissipation $\varepsilon_{t} := -\bm \tau_t : \bar{\bm S}$ represents the instantaneous power drained from the resolved scale into subgrid scales, and $\varepsilon_{\mu} = -\bm \tau : \bar{\bm S}$ represents the viscous dissipation of the resolved scales. The essential idea is to quantify the effects of the subgrid-scale model by comparing $\langle \varepsilon_{t} \rangle$ with $\langle \varepsilon_{\mu} \rangle$. In practice, a surrogate for the subgrid activity parameter is introduced by comparing the eddy viscosity to the physical viscosity \cite{Celik2005}, that is, $s := \langle\nu_t\rangle/(\langle\nu_t\rangle + \nu)$. By definition, $0 \leq s < 1$. Small values of $s$ indicate that the resolved scales are responsible for the majority of the energy cascade, whereas larger values of $s$ imply that the subgrid-scale model is handling a substantial portion of the dissipation. It is recommended that $s<0.3$ correspond to LES resolutions for which both modeling and discretization errors remain below $1\%$ \cite{Broglia2003,Geurts2002}, indicating a well-resolved turbulence simulation. An alternative index was proposed by Scillitoe et al. \cite{Scillitoe2015} and is defined as $\mathrm{IQ}_{\nu} := [ 1+\alpha_{\nu}(\langle \nu_{t}\rangle/\nu)^n]^{-1}$, where the parameters are calibrated as $\alpha_{\nu}=0.05$ and $n=0.53$. With this definition, $\mathrm{IQ}_{\nu}=0.8$ indicates a well-resolved LES result, and values of $0.95$ and above correspond to DNS-like results. This index has since been adopted in several studies to evaluate the quality of blood pump simulations \cite{Liu2022,Torner2020}.

Figure \ref{fig:contours_LES_quality_C2} shows the contours of the subgrid activity parameter and the index $\mathrm{IQ}_{\nu}$ on two representative planes at $y=0.006$ and $z=0.000$ for all three meshes under Condition 2. The two finer meshes exhibit consistently high LES quality, with only minor differences observed between them. Their subgrid activity parameters remain below $0.1$ over most regions, and they increase moderately to values between $0.1$ and $0.3$ in the rotor wake and throat of the diffuser. Their $\mathrm{IQ}_{\nu}$ values are also predominantly large, exceeding 0.98 in most regions. In contrast, Mesh 1 exhibits noticeably higher subgrid activity values, ranging between 0.2 and 0.6, in the rotor wake and diffuser regions. It also exhibits lower $\mathrm{IQ}_{\nu}$ values (between 0.96 and 0.98) compared with the other two meshes. These trends indicate Mesh 1 lacks sufficient resolution in key flow regions. Figure \ref{fig:contours_LES_quality_C5} presents the LES quality contours under Condition 5, revealing trends similar to those observed for Condition 2. Meshes 2 and 3 continue to demonstrate high LES quality according to both metrics. For these meshes, the subgrid activity parameters remain below 0.25 over most regions, and the corresponding $\mathrm{IQ}_{\nu}$ values exceed 0.98. By comparison, Mesh 1 exhibits elevated subgrid activity, with values between 0.2 and 0.6 over most of the region, corresponding to lower $\mathrm{IQ}_{\nu}$ values between 0.95 and 0.98. These results further indicate that Mesh 1 is close to the minimum grid resolution necessary for capturing the dominant flow features in critical pump regions.

\section{Local turbulent structures}
In blood pump applications, local turbulent features are of particular relevance, as regions of concentrated vortical activity and elevated TKE may be associated with increased risk of blood damage. We examine the large-scale vortical structures that govern momentum transport and the velocity energy spectra that characterize the transfer of turbulent energy across scales. Together, these analyses provide a more physically informative characterization of the blood pump flow. 

\begin{figure}[htbp]
  \centering
  \begin{minipage}[t]{0.9\textwidth}
    \centering
    \includegraphics[width=1\linewidth, trim=50 230 50 240, clip]{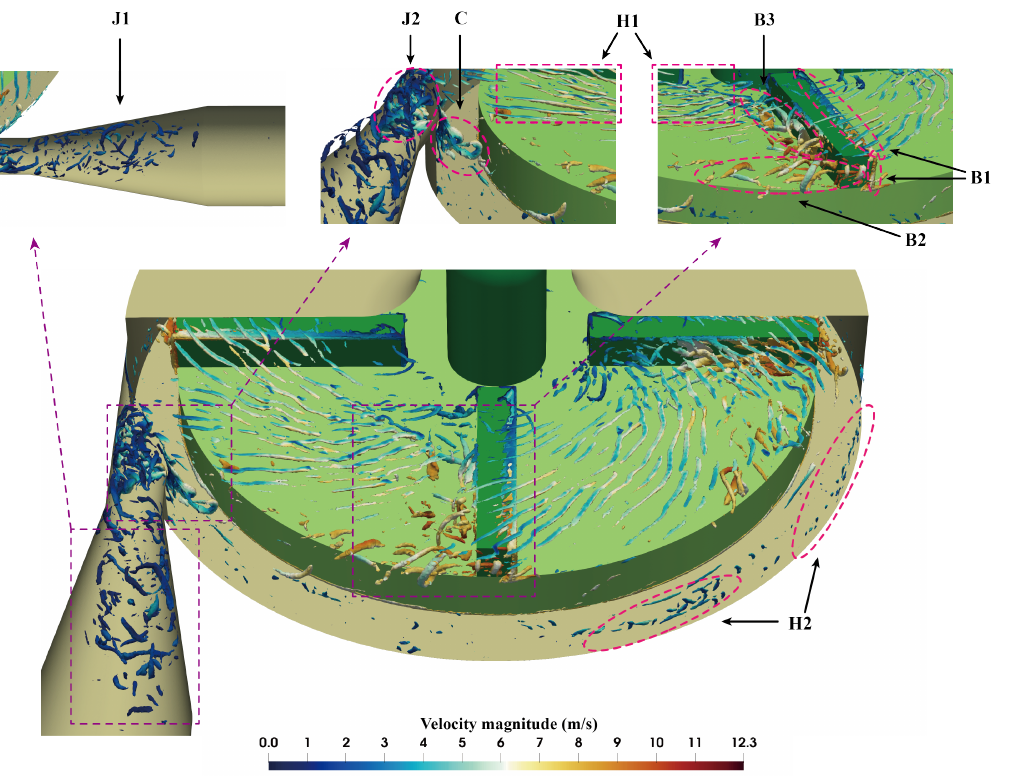}
    \caption{Vortical structures under Condition 2, isosurface of $\mathcal Q= 1.2 \times 10^7~\mathrm{s^{-2}}$.}
    \label{fig:vortex_80M_C2}
  \end{minipage}

  \begin{minipage}[t]{0.9\textwidth}
    \centering
    \includegraphics[width=1\linewidth, trim=50 270 50 250, clip]{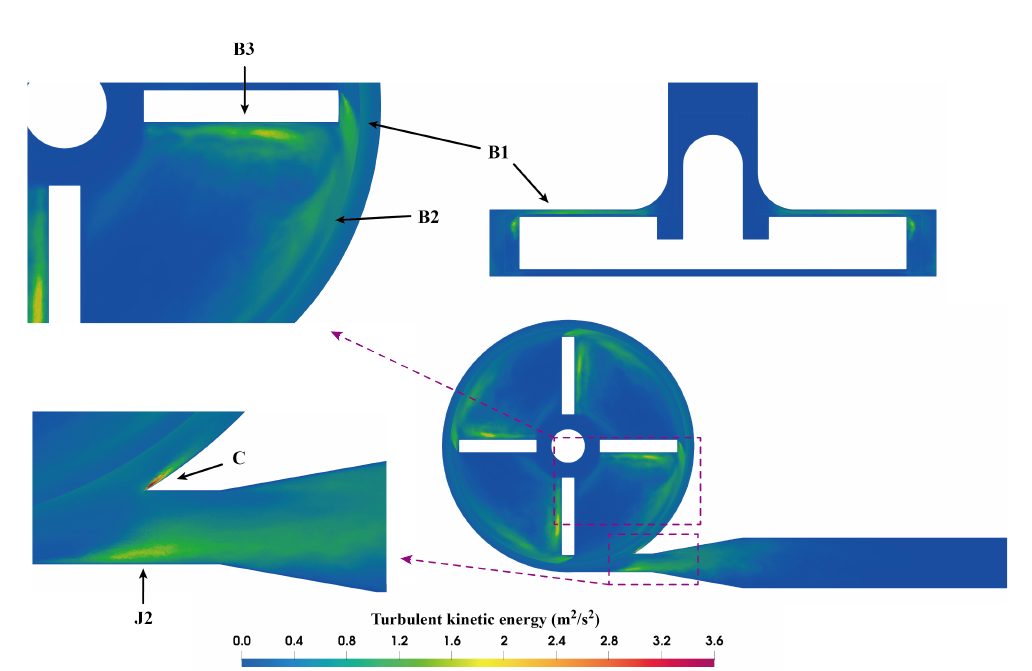}
    \caption{TKE contour under Condition 2.}
    \label{fig:TKE_80M_C2}
  \end{minipage}
\end{figure}

\begin{figure}[htbp]
  \centering
  \begin{minipage}[t]{0.9\textwidth}
    \centering
    \includegraphics[width=1\linewidth, trim=50 230 50 240, clip]{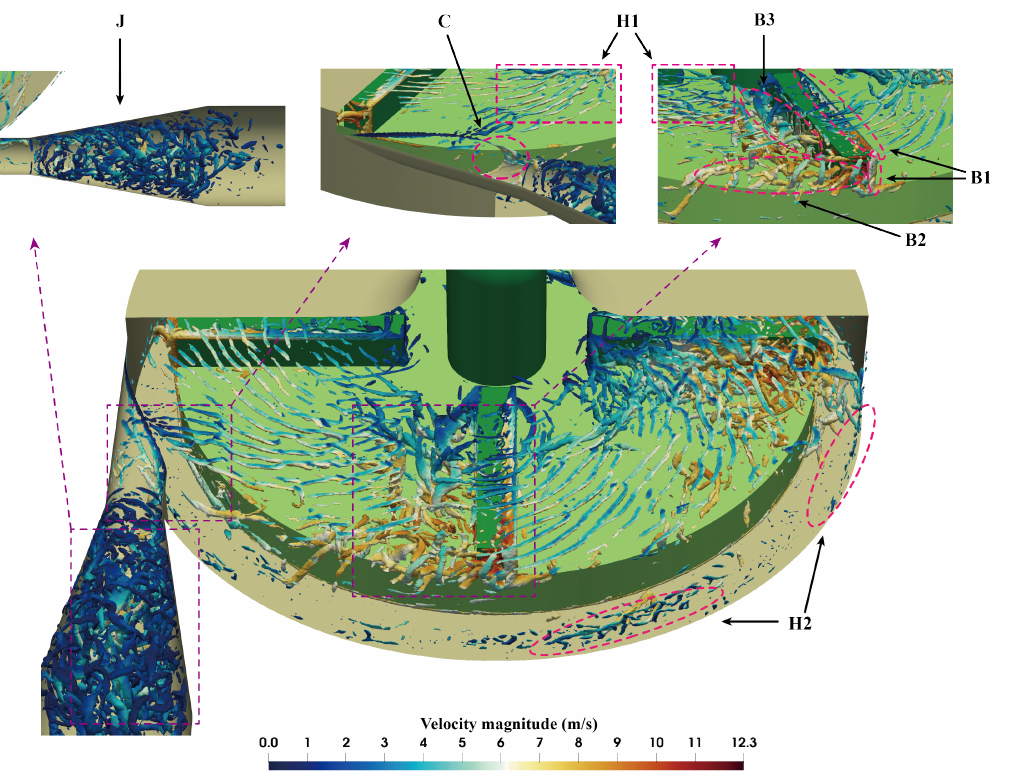}
    \caption{Vortical structures under Condition 5, isosurface of $\mathcal Q = 1.2 \times 10^7~\mathrm{s^{-2}}$.}
    \label{fig:vortex_80M_C5}
  \end{minipage}

  \begin{minipage}[t]{0.9\textwidth}
    \centering
    \includegraphics[width=1\linewidth, trim=50 270 50 250, clip]{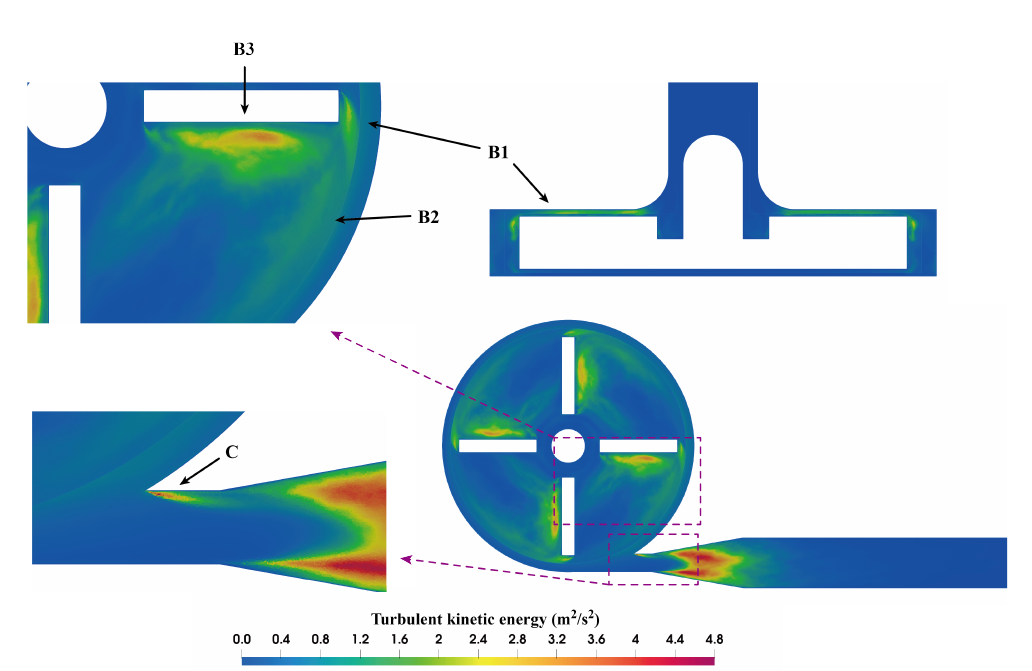}
    \caption{TKE contour under Condition 5.}
    \label{fig:TKE_80M_C5}
  \end{minipage}
\end{figure}

To visualize and analyze the vortical structures within the pump, the Q-criterion \cite{Hunt1988} is employed to detect and characterize the location and topology of large-scale vortical structures. It is defined as $\mathcal Q := \left(\bar{W}_{ij}\bar{W}_{ij} - \bar{S}_{ij}\bar{S}_{ij}\right)/2$, where $\bar{W}_{ij} :=(\partial\bar{u}_i/\partial x_j-\partial\bar{u}_j/\partial x_i)/2$ denote the antisymmetric part of the resolved velocity gradient. Figure \ref{fig:vortex_80M_C2} presents the instantaneous vortical structures under Condition 2, visualized using an isosurface of $\mathcal Q = 1.2 \times 10^7~\mathrm{s^{-2}}$, obtained with Mesh 3 at the end of the 35th rotor revolution. Strong vortical structures originate at the leading edges of the rotor blades (B1). These tip leading-edge vortices subsequently evolve into wake vortices (B2), where the locally elevated velocities induce the formation of twisted vortex tubes. Trailing-edge vortices (B3), extending from the blade root to the tip, are generated as the flow rapidly fills the low-pressure region behind the blade during its passage through the fluid. In addition, numerous narrow, stripe-like vortex tubes are observed adjacent to the upper housing wall (H1) within each quadrant. These structures originate from the leading-edge vortices above the blade and are subsequently compressed by the narrow gap between the blade top surface and the housing wall. The flow in this region is dominated by intense shear, which prevents the vortex tubes labeled by (B1) and (H1) from merging into a connected vortical structure. Several small, isolated vortices are present adjacent to the lateral housing (H2). Jet-induced vortices form in the throat region (J2) and gradually dissipate downstream in the diffuser (J1). The spatial distribution of the vortices (J1) is consistent with the jet direction in the diffuser, as shown in Figure \ref{fig:mean_velo_20r_30r}. Furthermore, a spiral near-wall vortex is observed adjacent to the cutwater (C), developing along the lateral housing as a branch from the throat vortex (J2).

Figure \ref{fig:TKE_80M_C2} shows the TKE contours for Condition 2 obtained with Mesh 3 on the sampling plane. It highlights regions of elevated turbulence intensity, primarily located in the vicinity of the blade tips, above the blade top surfaces, downstream of the blade trailing edges, within the blade wakes, and throughout the throat and diffuser. Notably, the near-wall vortex (C) is associated with a region of exceptionally high TKE adjacent to the cutwater and along the lateral housing wall, despite the local velocity magnitude being only moderate. This observation suggests that the enhanced turbulence production in this region may be associated with vortical activity rather than with high mean velocities. Overall, the spatial correspondence between regions of high TKE and intense vortical structures underscores the close link between vortex dynamics and turbulence generation in the pump flow.

Figure \ref{fig:vortex_80M_C5} presents the instantaneous vortical structures under Condition 5 at the same time instance, visualized with the same value of $\mathcal Q$. Overall, the spatial organization of vortical structures within the pump remains qualitatively similar to those observed in Condition 2, including the presence of leading-edge vortices (B1), wake vortices (B2), trailing-edge vortices (B3), and near-wall vortices on the housing (H1 and H2). Despite the similarities, the vortical structures here are characterized by a higher density of vortex tubes, increased spatial extent, and more pronounced twisting and fragmentation, indicating enhanced vortex stretching and stronger interactions across multiple scales. In particular, the leading-edge vortices (B1) are associated with higher local velocity magnitudes, while the wake vortices (B2) display enhanced twisting and stronger interaction with the surrounding flow, suggesting pronounced vortex stretching and turbulent mixing. The trailing-edge vortices (B3) also extend over a larger spatial region, reflecting intensified flow separation and wake development behind the blades. With respect to the jet flow, the throat region exhibits an almost laminar flow pattern, which suppresses the formation of jet-induced vortices within the throat. As a result, coherent jet-induced vortical structures develop primarily in the diffuser region (J), rather than in the throat as observed under Condition 2. The near-wall vortex (C) adjacent to the cutwater is less pronounced under Condition 5 and exhibits a tongue-like structure along the throat wall.

The TKE contours for Condition 5, shown in Figure \ref{fig:TKE_80M_C5}, also exhibit a close correspondence between regions of elevated TKE and the distribution of vortical structures. Compared with Condition 2, the overall TKE level is markedly higher, indicating a more energetic turbulent flow regime under this operating condition. Variations in the TKE distribution close to the cutwater further reflect the altered characteristics of the near-wall vortex (C) under this condition. Within the diffuser region, two distinct high-TKE zones are evident, indicating oscillatory behavior of the jet accompanied by intermittent switching between preferred flow paths.

\begin{figure}[htbp]
  \centering
  \begin{tabular}{ccc}
    & \quad\quad {Point A}&
    \quad\quad {Point B} \\[0.5em]
    \raisebox{1.15\height}{\rotatebox{90}{Condition 2}} &
    \includegraphics[width=0.4\linewidth, trim=0 210 30 230, clip]{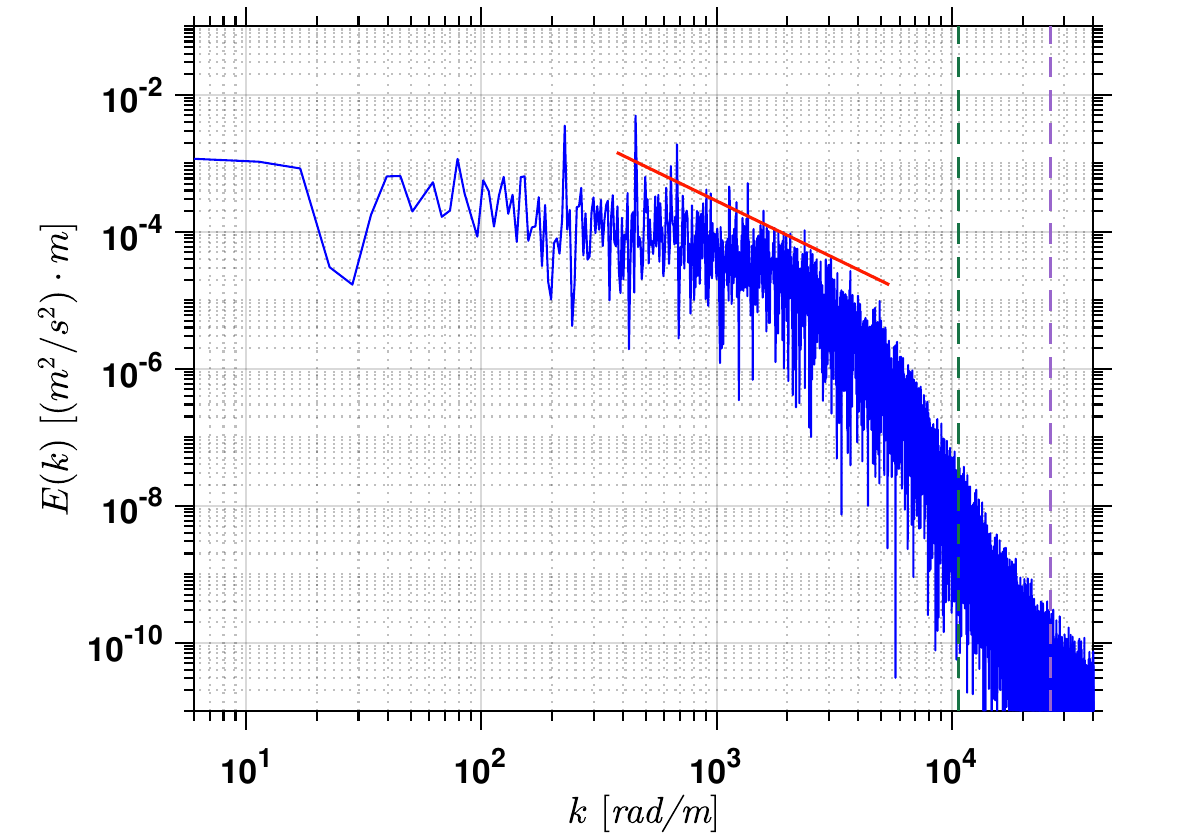} &
    \includegraphics[width=0.4\linewidth, trim=0 210 30 230, clip]{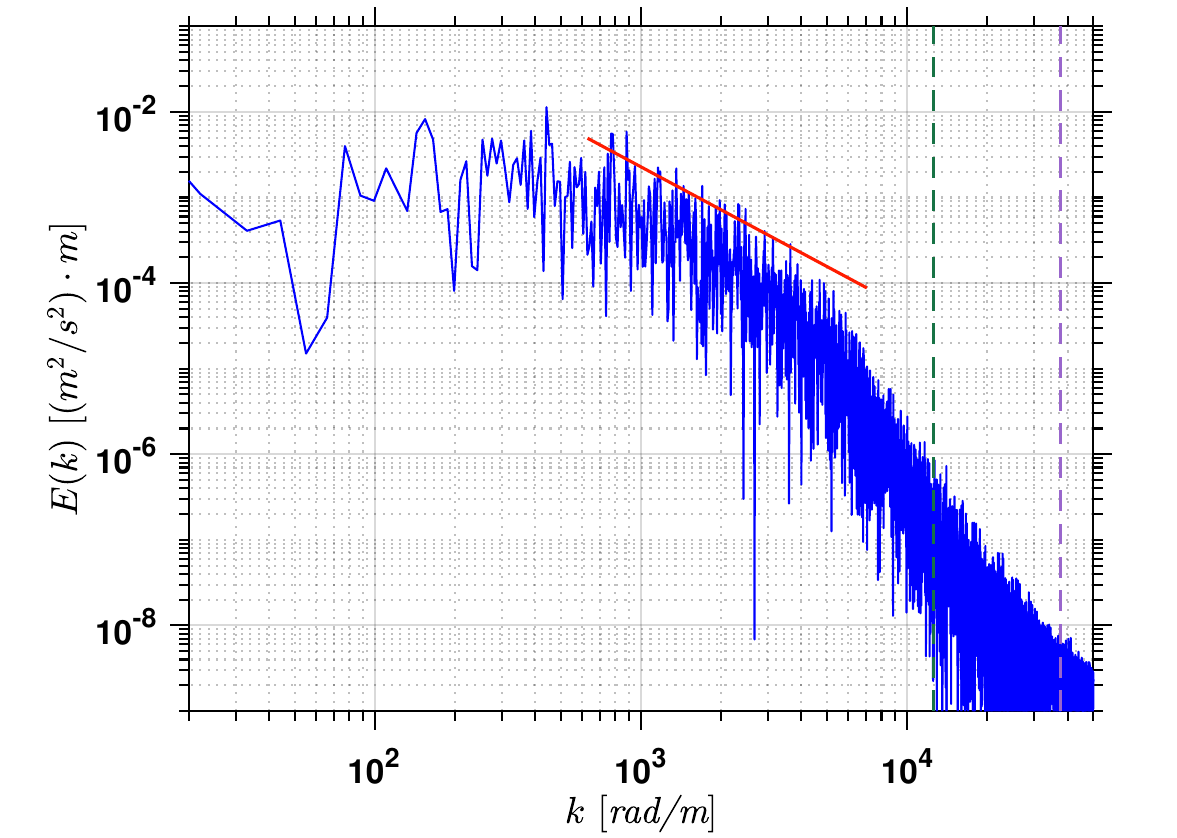} \\
    \raisebox{1.15\height}{\rotatebox{90}{Condition 5}} &
    \includegraphics[width=0.4\linewidth, trim=0 190 30 230, clip]{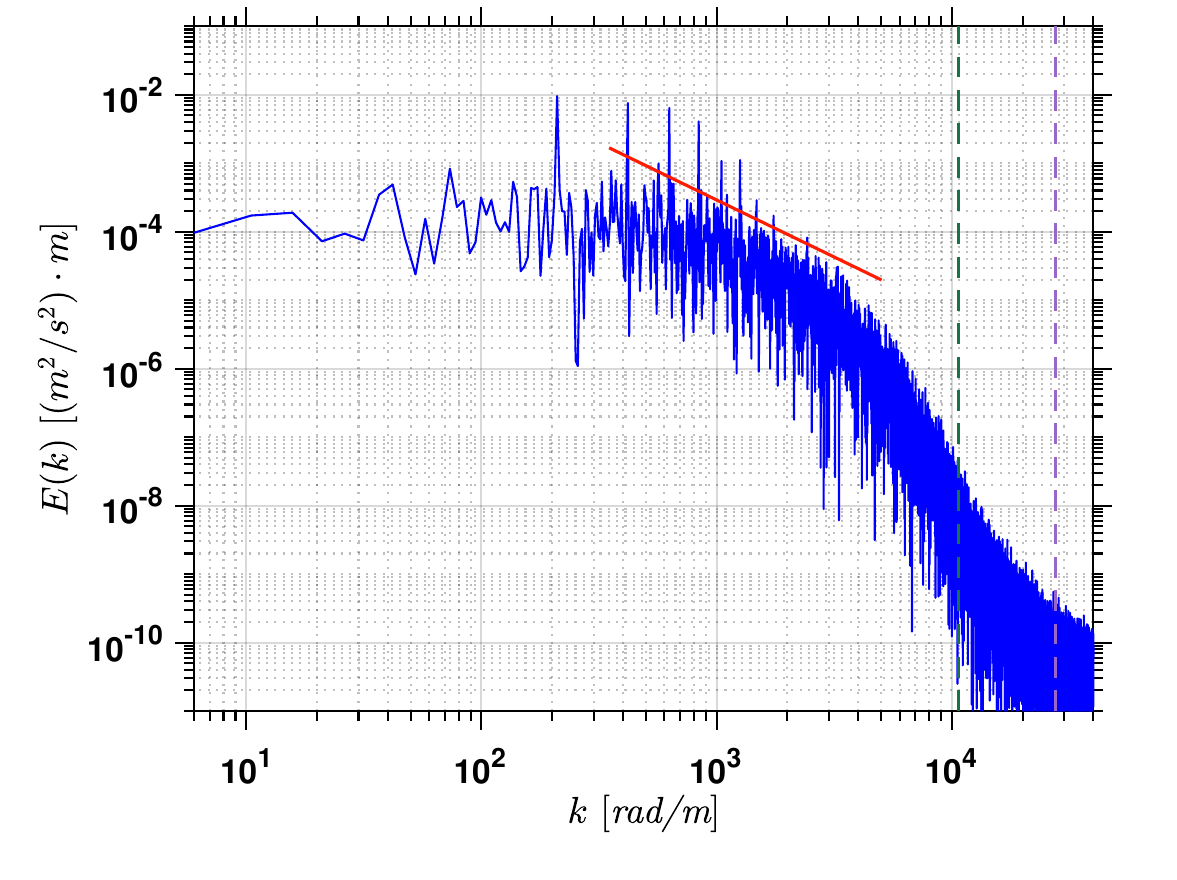} &
    \includegraphics[width=0.4\linewidth, trim=0 190 30 230, clip]{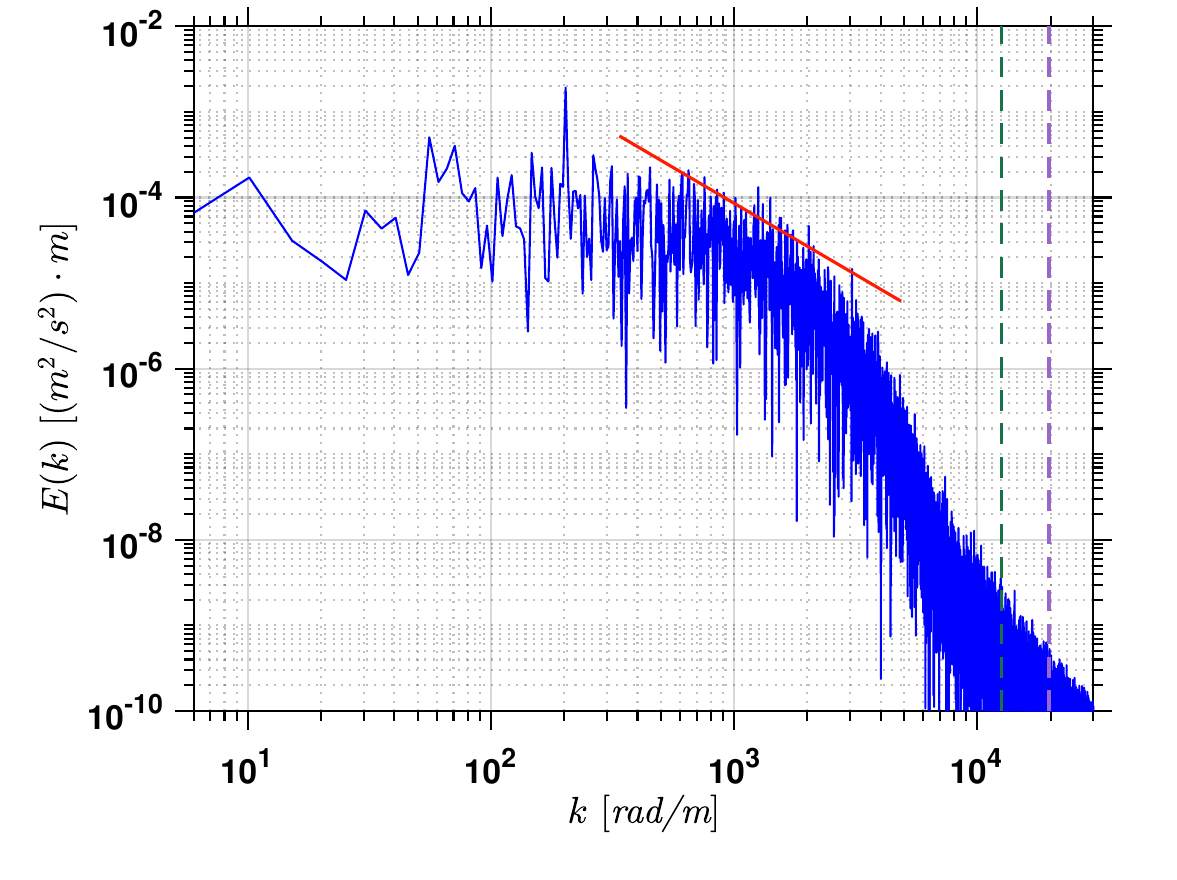} \\
  \end{tabular}
  \includegraphics[width=0.45\linewidth, trim=0 0 0 0, clip]{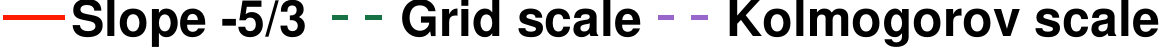}
  \caption{Velocity energy spectra at Points A and B.}
  \label{fig:energy_spectra_80M}
\end{figure}

To further investigate the turbulent characteristics within the pump, velocity energy spectra are evaluated at two representative locations on the sampling plane: point A in the blade passage region, located at $r=20\sqrt{2}~\mathrm{mm}$, and point B at the diffuser entrance along the centerline, as illustrated in Figure \ref{fig:pump_geometry}. The spectra are inferred from single-point velocity time series using Taylor's frozen turbulence hypothesis. Figure \ref{fig:energy_spectra_80M} presents the spectra obtained with Mesh 3. Both locations exhibit spectral ranges with slopes close to the classical $-5/3$ scaling associated with the inertial subrange. At higher wavenumbers, the spectra enter the dissipation range, where the spectral energy decreases more rapidly. Similar dissipation-range behavior has been reported in previous studies of axial blood pumps \cite{Torner2020}, suggesting that comparable small-scale dynamics are present in the current pump configuration. In all cases, the characteristic scales associated with the inertial and dissipation ranges are larger than the grid scale, while the grid scale remains larger than the Kolmogorov length scale estimated by $(\nu^{3} / \langle \varepsilon_{\mu} \rangle)^{1/4}$. This scale separation indicates that the inertial subrange dynamics are adequately resolved, while the smallest dissipative scales remain unresolved, consistent with the intended resolution characteristics of LES. At low wave numbers ($k_1 < 1 \times 10^3~\mathrm{rad/m}$), mild oscillations are observed in the spectra, likely reflecting large-scale turbulent intermittency modulated by the periodic blade-passing dynamics of the pump. Notably, although turbulence onset in pumps is commonly considered to occur when the pump Reynolds number exceeds $10^6$ \cite{WuLetter}, pronounced turbulent flow features are observed here at lower pump Reynolds numbers, as evidenced by the TKE distributions and energy spectra.

\section{Conclusion}
This study presents a systematic validation and comparative assessment of modeling strategies for blood pumps, against FDA benchmark experimental data. Direct comparison with experiments demonstrates that LES combined with SI provides reliable predictions of the flow field. In the blade passage, LES yields consistent results even on the coarsest mesh. In the diffuser, LES also predicts mean velocity profiles that are in close agreement with the experimental data. In this region, the lower local critical Reynolds number implies an increased tendency toward transition to turbulence, making scale-resolving approaches especially well-suited. In contrast, the RANS-MRF approach, which is widely adopted in the literature \cite{Han2022,Miccoli2024,Mohammadi2022,Sahebi-KuzehKanan2025}, exhibits noticeable discrepancies and nonphysical oscillations in the blade passage. Both RANS and URANS approaches exhibit substantial deviations from the experimental data or the LES results in critical regions. Further analysis reveals that the turbulence in the blood pump arises from a complex interplay between blade-induced vortices, wake structures, near-wall vortical motions, and jet flow, whose spatial organization is closely correlated with regions of elevated TKE. This correspondence highlights the central role of vortex dynamics in turbulence generation within the pump. The comparison also reveals the impact of operating conditions on the variations in vortex stretching and jet development. Complementary spectral analysis further confirms that a substantial portion of the turbulence spectrum is directly resolved by the present LES, with clear scale separation.

Collectively, these results reinforce the suitability of LES for capturing the turbulent mechanisms. Since most previous numerical studies of blood pumps relied primarily on RANS-based approaches, the present analysis demonstrates that such methods may be insufficient to capture the pronounced unsteadiness and rich multiscale flow structures observed in the pump. By directly resolving a substantial portion of the turbulent fluctuations, transient scale-resolving LES offers superior reliability in capturing the underlying flow physics, which is expected to play a critical role in improving the fidelity of blood damage predictions \cite{Kameneva2004, Torner2019}.

To provide practical guidance for future LES studies of VADs and related medical devices, a series of calculations were carried out using three levels of spatial resolution, and the simulation quality was assessed using three complementary evaluation metrics. The results indicate that Mesh 1, with approximately 10 million elements, represents a marginal resolution threshold for the FDA benchmark pump. Coarser meshes are likely insufficient to adequately resolve the dominant flow features in critical regions with strong turbulence activity. In contrast, the mesh with 80 million elements resolves a substantially larger fraction of the TKE. For the prediction of mean flow quantities, however, the improvements relative to the intermediate mesh are modest, indicating that this level of resolution may exceed what is strictly necessary for reliable mean-flow characterization in the present configuration. However, such a level of spatial resolution may still be required for studies focused on blood damage, where accurately resolving small-scale flow structures and elevated shear regions can be of particular importance. Overall, these findings delineate a practical resolution range and offer quantitative guidance for balancing predictive accuracy against computational cost in future investigations.

Limitations of the present study should be acknowledged. First, a Newtonian blood model was employed in accordance with the benchmark setting. Previous studies have shown that non-Newtonian effects may influence flow structures and hemolysis predictions in blood pumps \cite{Good2020}. Second, hemolysis was not explicitly evaluated in this work. While the resolved turbulent features captured by LES are expected to influence hemolytic behavior differently from RANS-based predictions, quantifying this effect requires dedicated hemolysis modeling and validation. Future work will therefore focus on extending the present framework to improve the assessment of both hydrodynamic performance and blood damage in VADs.

\section*{Acknowledgements}
This work is supported by the National Natural Science Foundation of China [Grant Numbers 12472201,12172160], Shenzhen Science and Technology Program [Grant Number JCYJ20220818100600002], Southern University of Science and Technology [Grant Number Y01326127], and the Department of Science and Technology of Guangdong Province [2021QN020642]. Computational resources are provided by the Center for Computational Science and Engineering at the Southern University of Science and Technology.

\bibliographystyle{unsrt}
\bibliography{fsi-n-cfd}

\end{document}